\documentclass[pdflatex,sn-nature]{sn-jnl}
\usepackage{hyperref}
\usepackage[T1]{fontenc}
\usepackage[utf8]{inputenc}
\usepackage{graphicx}
\usepackage{booktabs}
\usepackage{amsmath,amssymb}
\usepackage{multirow}
\usepackage{xcolor}
\usepackage{manyfoot}
\usepackage{array}
\usepackage{url}
\usepackage{enumitem}

\graphicspath{{figures/}{supplementary/}}

\newcommand{\asirn}{a-Si$_3$N$_4$}
\newcommand{\iprr}{I_{\mathrm{atom}}}

\begin{document}

\title{State aligned graph wavelet fingerprints of electron trapping in amorphous Si$_3$N$_4$}

\author[1,2]{\fnm{Jonathon} \sur{Cottom}}
\author[1,2]{\fnm{Emilia} \sur{Olsson}}

\affil[1]{\orgdiv{Institute for Theoretical Physics}, \orgname{University of Amsterdam}, \orgaddress{\city{Science Park 904, 1098 XH, Amsterdam}, \country{the Netherlands}}}
\affil[2]{\orgname{Advanced Research Center for Nanolithography}, \orgaddress{\city{Science Park 106, 1098 XG Amsterdam}, \country{the Netherlands}}}

\abstract{Classifying charge traps in amorphous materials requires the electronic state and the network response to be resolved together. We develop state-aligned graph-wavelet fingerprints and apply them to electron trapping in \asirn{}, where intrinsic-polaronic trapping, coordination-driven trapping, and charge-driven conversion coexist within one ensemble. Across 415 pairs of relaxed neutral and negatively charged cells, alignment of the trap state before and after localisation connects electronic identity to multiscale structural fingerprints. Charge-driven conversion produces substantial reorganisation across graph scales and increases the fingerprint amplitude. The fingerprint change identifies the transformation each site undergoes, while the occupation-induced level shift tracks localisation gain and graph deformation tracks stabilisation. Charge-induced defect formation dominates the deepest spectral-trap decile: 69\% form at sites whose Si--N bond breaks on capture, although this route accounts for 44.2\% of the classified traps. The fingerprints unify electronic identity, multiscale network reorganisation and energetic stabilisation in a quantitative framework for resolving charge trapping in amorphous materials.}


\maketitle

\section{Introduction}

Few disordered dielectrics provide a more stringent test of charge-trap classification than amorphous silicon nitride.\cite{robertson1981,robertson1984,warren1993} Charge trapping in nitride-containing stacks underpins non-volatile memory operation, making the underlying trapping mechanisms central to threshold-voltage stability and long-term reliability. Resolving those mechanisms requires relating ensemble measurements to diverse local environments in an amorphous network without a single stable defect archetype.\cite{robertson2006highk,gritsenko2016,cottom2024,Wilhelmer2026Review} In \asirn{}, that ambiguity is heightened rather than resolved by the literature: intrinsic-like K-centre physics, paramagnetic defect states, dangling-bond models, negative-$U$ behaviour, and charge-driven bond reorganisation have all been invoked to explain trapping in closely related local environments.\cite{robertson1983,robertson1984,robertson1994,warren1991dangling,warren1993,krick1988,kamigaki1990,petersen2006,gritsenko2016,vedula2013,kang2018,choi2024,cottom2024,Huckmann_22} The central question is therefore not which single defect picture is universally correct, but which trapping modes coexist within one disordered ensemble, which electronic states realise them, and how the surrounding network reorganises when charge is captured.

Charge-trap classification faces two coupled problems: establishing electronic-state identity across charge capture and resolving the accompanying network reorganisation. Between the neutral and charged systems, frontier ordering can change, localised states can shift in energy, and band-edge windows can remain similar while the trap-bearing state shifts substantially. Energetic proximity, frontier rank and final-state localisation alone do not establish correspondence between the precursor and trapped states. The structural problem is to determine how the trapping environment forms and reorganises. In \asirn{}, charge capture can further localise an intrinsic-polaronic precursor, stabilise a pre-existing defect or create an undercoordinated trapping centre through bond rearrangement.\cite{petersen2006,vedula2013,kang2018,choi2024,cottom2024,Huckmann_22,Wilhelmer2024Polaron} Electron-spin-resonance and optical measurements establish carrier localisation and metastable defect formation across amorphous oxides, nitrides and chalcogenides.\cite{griscom1992sth,hosono2001disorder,warren1993,bishop1977metastable} First-principles studies connect carrier trapping and defect-state stability to network relaxation and bond rearrangements, establishing their coupled electronic and structural character.\cite{boero1997eprime,binder2011trapping,guo2016oxygen,konstantinou2022gst_trapping,Huckmann_22} Charge trapping must therefore be characterised through aligned precursor and trapped states together with the multiscale response of the surrounding atomic network. This requires a unified framework that connects electronic-state identity, chemically interpretable network reorganisation and energetic stabilisation.

Graph-wavelet fingerprints provide a multiscale, multimetric description of the trapping environment. Once the trap-bearing state is identified, diffusion-wavelet channels resolve the surrounding network across local and intermediate graph scales.\cite{CoifmanMaggioni2006,CoifmanLafon2006,Hammond2011,Shuman2013,Chung1997,Ortega2018} The fingerprints distinguish precursor amplitude from charge-induced reorganisation, resolving the magnitude, chemical character and scale dependence of the response across the surrounding neighbourhood.\cite{Mallat2012,Sun2009,Gama2019DiffusionScattering,Gao2019GeometricScattering} By linking the local chemistry to collective network relaxation, these fingerprints identify the coupled electronic-structural transformations that create and stabilise charge traps.

Here we apply this framework to electron trapping in \asirn{}, combining alignment of the trap state before and after localisation with graph-wavelet fingerprinting of the responding network. The paired analysis resolves changes in localisation and electronic-centre position between the neutral precursor and relaxed trap. Graph-wavelet channels quantify the accompanying network response across dimensionless graph scales $t=0.5$--$8$. The same graph treatment is applied separately to local potential and spin-polarisation fields.

Within the \asirn{} ensemble, intrinsic-like, defect-driven and conversion-driven trapping coexist. The paired graph readouts distinguish traps with similar precursor amplitudes but different responses to charge capture. Here we show that charge-induced Si--N bond breaking at initially four-coordinate sites creates the undercoordinated centres that dominate the deepest electron traps. We therefore proceed from state identity to multiscale response and then to chemical interpretation. We first establish the workflow and a representative alignment, then derive the graph-wavelet fingerprints, ground the analysis against localisation gain, local ionic relaxation and stabilisation energy, and connect these responses to the trapping routes.

\section{Results}

\begin{figure}[ht]
\centering
\includegraphics[width=0.95\textwidth]{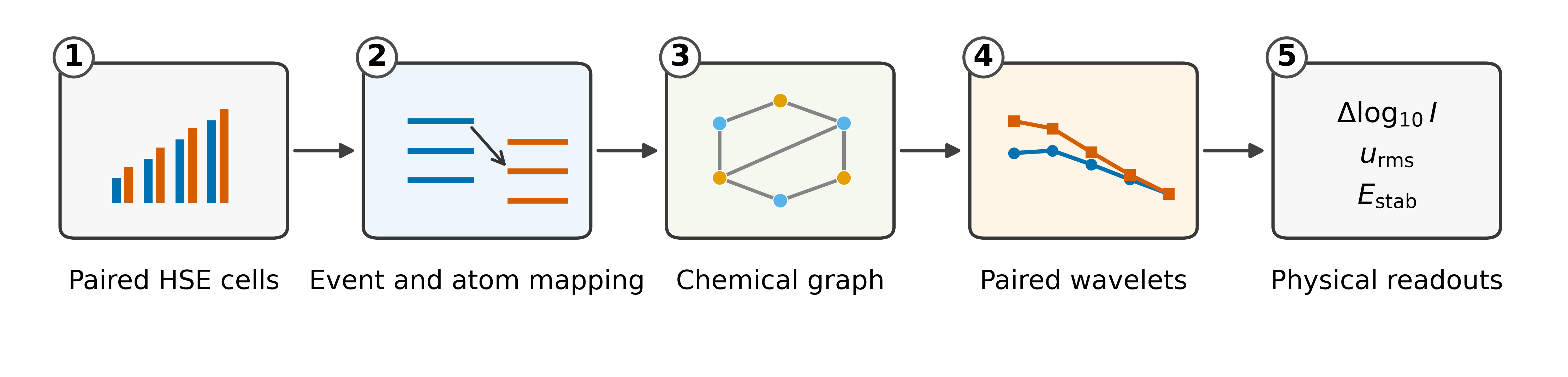}
\caption{\textbf{State-aligned graph-wavelet workflow.} Alignment of the precursor and trapped states establishes the electronic comparison. Diffusion-wavelet channels resolve the paired network response across $t=0.5$--$8$, connecting localisation, electronic-centre relocation, ionic relaxation, corrected stabilisation and the trapping routes.}
\label{fig:workflow}
\end{figure}

\subsection{State alignment resolves localisation and relocation}
The dataset comprises 415 pairs of relaxed neutral and negatively charged cells (Supplementary Table~S1). Alignment of the trap state before and after localisation connects each electronic response to the accompanying network reorganisation. The paired analysis resolves changes in localisation and electronic-centre position alongside the diffusion-wavelet response at $t=0.5$, $1$, $2$, $4$ and $8$ (Fig.~\ref{fig:workflow}). Precursor amplitude and charge-induced fingerprint change provide the principal structural readouts.

\begin{figure}[t]
\centering
\includegraphics[width=\textwidth]{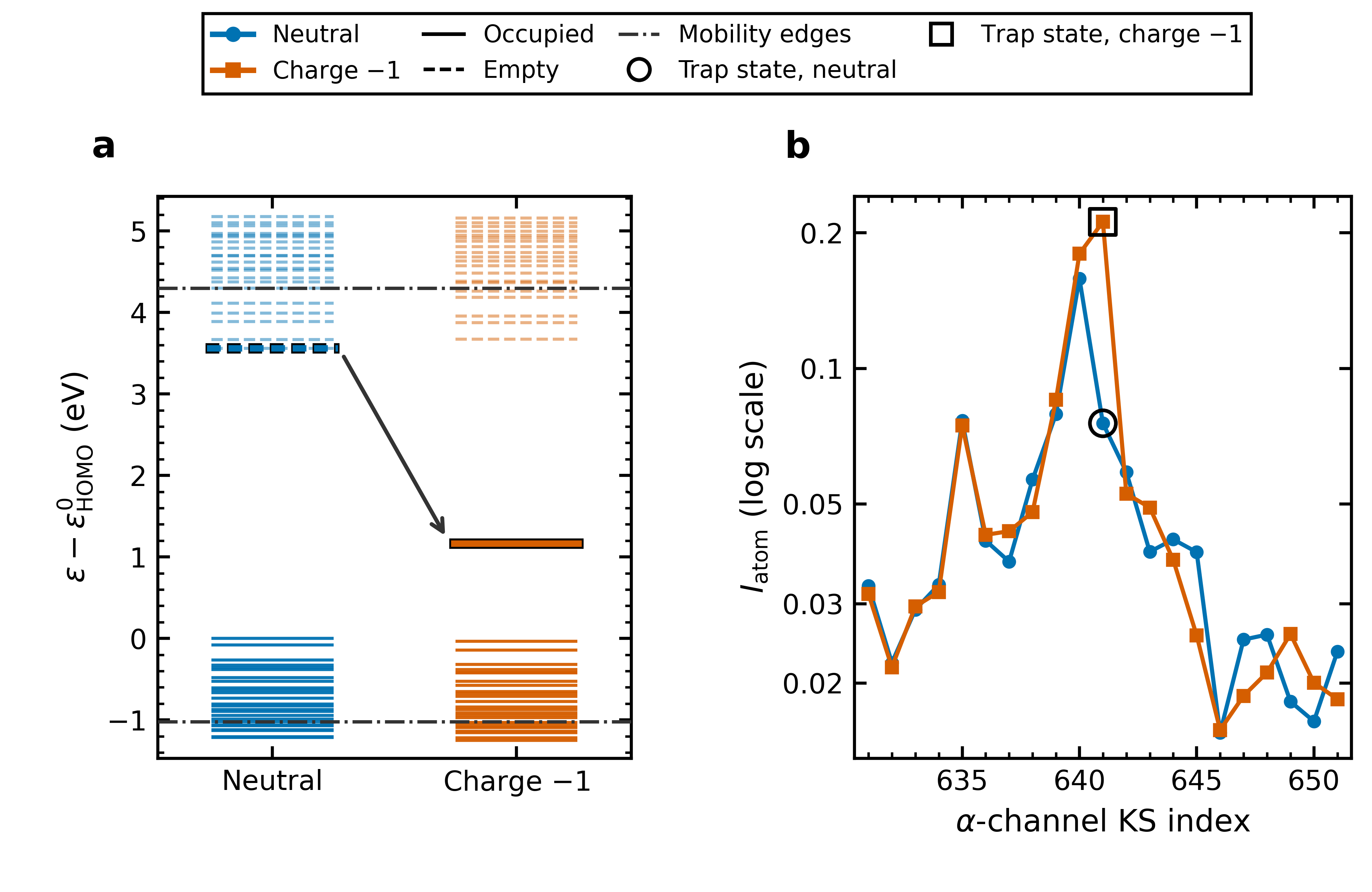}
\caption{\textbf{Representative state alignment for an intrinsic-like polaron trap.} \textbf{a,} In \texttt{box\_106}, the same $\alpha$/mo641 state moves from the neutral LUMO to the occupied charged-state HOMO. Levels are aligned through the N 2s reference and shown relative to the neutral $\alpha$ HOMO; solid and dashed levels denote occupied and empty states. The dash-dotted lines mark the neutral mobility edges at $-1.020$ and $+4.300$~eV, defined by $N_{\rm eff}^{\rm el}\geq56$. \textbf{b,} Atom-projected IPR for the aligned $\alpha$-channel frontier window, shown on a logarithmic axis. The highlighted state gains localisation on charging by $\Delta\log_{10}I_{\rm atom}=0.448$, with effective participation falling from 13.24 to 4.72 atoms.}
\label{fig:alignment}
\end{figure}

Figure~\ref{fig:alignment} shows a representative polaron trap (\texttt{box\_106}). The aligned $\alpha$/mo641 state moves from the neutral LUMO to the occupied charged-state HOMO on the common N 2s-aligned energy scale, referenced to the neutral mobility edges (Fig.~\ref{fig:alignment}a). Its effective participation falls from 13.24 to 4.72 atoms, giving $\Delta\log_{10}(\iprr)=0.448$ (Fig.~\ref{fig:alignment}b), while its atom-projected character remains Si-centred. Electron capture therefore further localises an already localised precursor. Neutral localisation profiles and route-specific spectral examples are shown in Supplementary Figs.~S3 and S4, respectively. Across the full ensemble, the selected trap level lies below the anion HOMO in 42 cells (Supplementary Fig.~S5). The final trapping atom lies outside the precursor participation region in four of the 412 classified cell pairs (Supplementary Tables~S1 and S2). The paired analysis thus retains both spectral reordering and electronic-centre relocation.

\subsection{Precursor environments and multiscale responses distinguish trapping routes}
The 412 classified traps fall into three trapping routes: polarons (Si$_4\!\rightarrow$Si$_4$, $n=165$), induced K centres (Si$_4\!\rightarrow$Si$_3$, $n=182$), and native K centres (Si$_3\!\rightarrow$Si$_3$, $n=65$). Three traps have no assigned route and are treated separately (Supplementary Table~S6). The multiscale graph response is summarised by the precursor amplitude $T^0$ and the relative fingerprint change $\eta=\|V^--V^0\|_2/\|V^0\|_2$, where $V$ contains all five scales and seven structural channels.

\begin{figure}[ht]
\centering
\includegraphics[width=\textwidth]{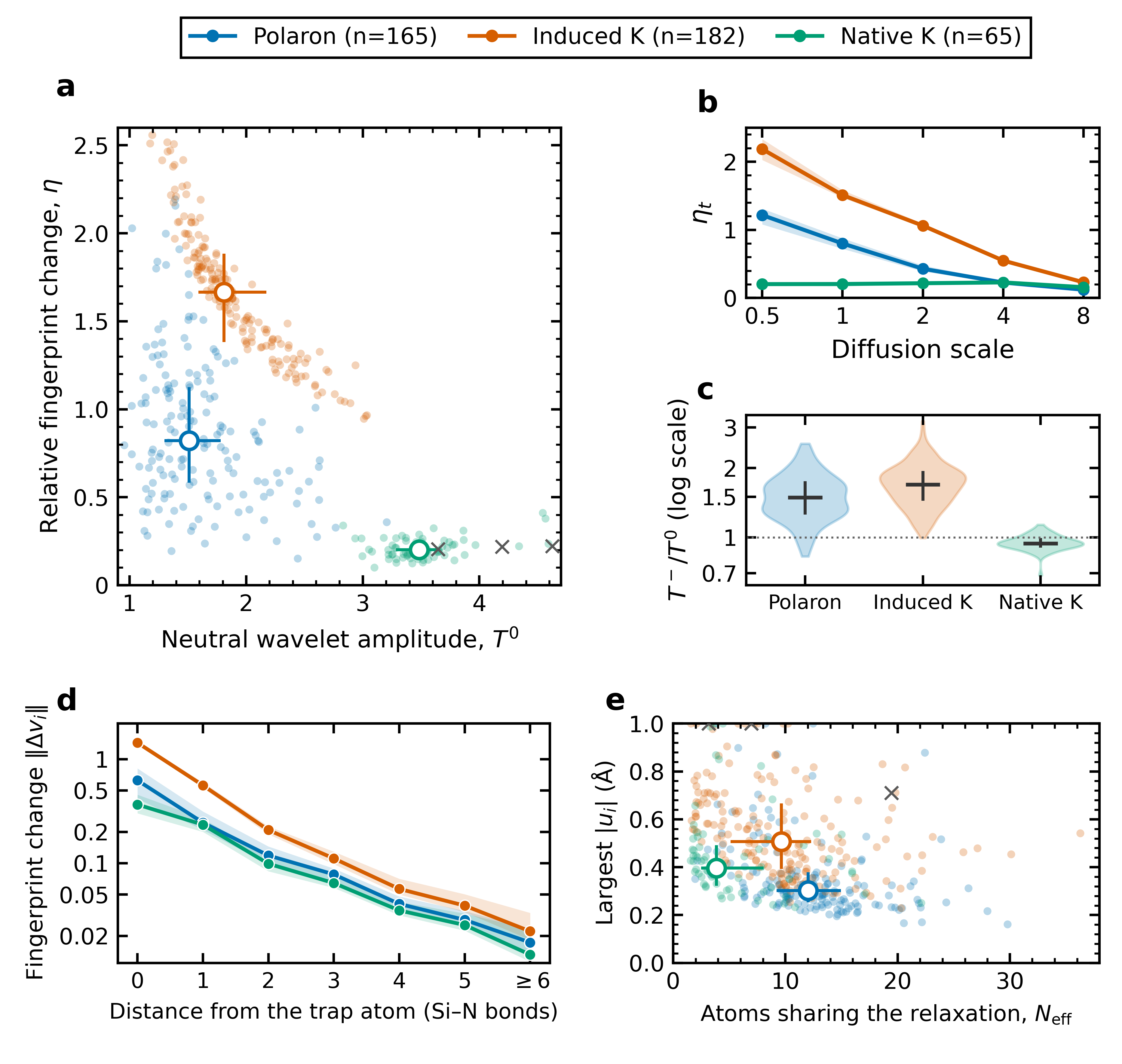}
\caption{\textbf{Graph-wavelet fingerprints separate neutral amplitude from charge-induced reorganisation.} \textbf{a,} Neutral amplitude $T^0$ and relative fingerprint change $\eta$. Open markers and bars show route medians and interquartile ranges; grey crosses denote the three unclassified cells. Five induced-K data points lie outside the displayed window; all enter the summaries. \textbf{b,} Median relative fingerprint change $\eta_t$ across scale, with 95\% whole-cell bootstrap intervals. \textbf{c,} Amplitude ratios $T^-/T^0$ on a logarithmic axis, with median and interquartile marks; unity denotes unchanged amplitude. \textbf{d,} Fingerprint change $\|V_i^--V_i^0\|_2$ against neutral-graph distance from the trap atom. Lines are route medians of per-pair shell medians, and bands are interquartile ranges; distances of six or more bonds are pooled. \textbf{e,} Maximum atomic displacement against effective displacement participation $N_{\rm eff}^{u}$, with route medians and interquartile ranges; displacements above 1~\AA{} are drawn at the frame. The spatial profile in \textbf{d} describes the extent of fingerprint change; \textbf{e} measures the physical relaxation.}
\label{fig:wavelets}
\end{figure}

Precursor amplitude and charge-induced reorganisation distinguish the three trapping routes (Fig.~\ref{fig:wavelets}a). The two readouts correctly classify 95.7\% of traps under cross-validation (Supplementary Table~S7). Precursor fingerprint vectors give a receiver-operating-characteristic area under the curve (AUC) of 0.83 for polarons versus induced K centres, and trapped fingerprint vectors give 0.87 for induced versus native K centres. The paired response vectors separate every route pair (AUC 1.00). The coordination contribution is evident in $\eta$: without these channels, polaron-induced-K separation falls to AUC 0.50. Unfiltered trap-atom channels give comparable route separation (Supplementary Table~S7). The fingerprints connect this chemical distinction to the scale and spatial extent of network reorganisation. The neutral amplitudes of polarons and induced K centres overlap, with medians of 1.51 and 1.81, compared with 3.48 for native K centres. Median $\eta$ is 0.821 for polarons, 1.665 for induced K centres, and 0.201 for native K centres. The paired fingerprints thus distinguish charge-driven formation of K centres from capture at pre-existing K centres.

The polaron and induced-K responses are largest at short diffusion scales, with median $\eta_t$ falling from 1.22 to 0.12 and from 2.19 to 0.23, respectively, whereas the native-K response remains near 0.2 (Fig.~\ref{fig:wavelets}b). The structural amplitude increases for polarons and induced K centres: median $T^-/T^0$ values are 1.491 and 1.696, compared with 0.942 for native K centres (Fig.~\ref{fig:wavelets}c). The induced-K fingerprint change also extends across several graph shells (Fig.~\ref{fig:wavelets}d). Its physical relaxation has a median effective participation of 9.6 atoms and a median maximum displacement of 0.51~\AA{}, compared with 12.0 atoms and 0.30~\AA{} for polarons and 3.8 atoms and 0.40~\AA{} for native K centres (Fig.~\ref{fig:wavelets}e). Broad descriptor change therefore coexists with moderate displacement participation.

\subsection{Electronic and network responses connect localisation, relaxation and stabilisation}
The three trapping routes span overlapping distributions of precursor localisation, ionic relaxation and stabilisation (Supplementary Table~S2). Precursor participation, local ionic RMS displacement and corrected stabilisation span 4--33 effective atoms, 0.12--0.30~\AA{} and 0.9--2.1~eV, respectively (5th--95th percentiles). Route-mean baselines give held-out $R^2=0.23$ for localisation gain and 0.13 for stabilisation (Supplementary Table~S3).

\begin{figure}[ht]
\centering
\includegraphics[width=\textwidth]{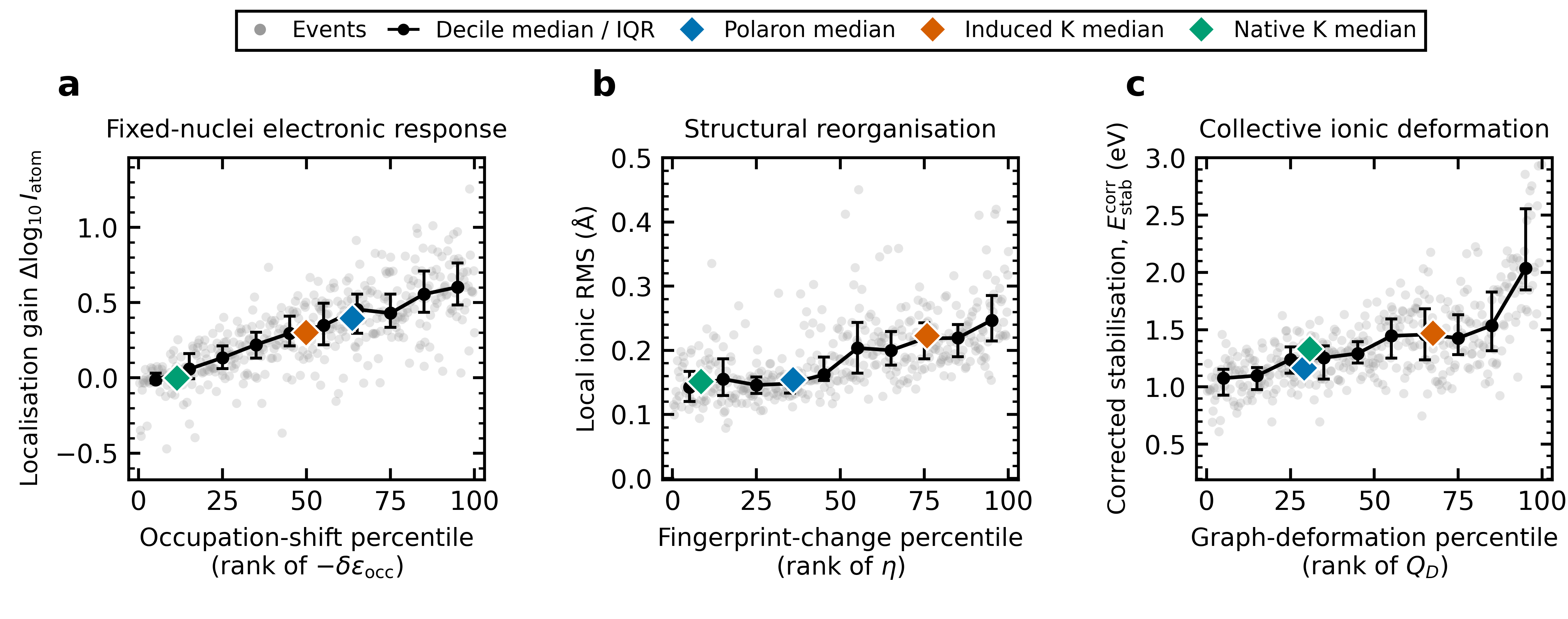}
\caption{\textbf{Physical grounding of the graph-wavelet response.} \textbf{a,} Electronic localisation gain after relaxation against the fixed-nuclei occupation response $-\delta\epsilon_{\rm occ}$. \textbf{b,} Local RMS displacement of the trap neighbourhood against charge-induced fingerprint change $\eta$. \textbf{c,} Corrected total-energy stabilisation against collective graph deformation $Q_D$. The horizontal axes are shown as within-ensemble percentiles, the black symbols and line show decile medians with interquartile bars, and the coloured markers show route medians. Three data points in \textbf{b} and four in \textbf{c} lie above the display limits; all enter the summaries.}
\label{fig:grounding}
\end{figure}

The fixed-nuclei occupation response $-\delta\epsilon_{\rm occ}$ tracks the net localisation gain on trapping (Spearman $\rho=0.765$, $n=412$; Fig.~\ref{fig:grounding}a), with within-route correlations of 0.49--0.74. It also tracks precursor participation ($\rho=0.83$; Supplementary Fig.~S10d). Structural fingerprint change tracks local ionic RMS displacement ($\rho=0.636$, $n=412$; Fig.~\ref{fig:grounding}b), with within-route correlations of 0.31--0.50. Its single-coordinate calibration gives $R^2=0.26$, compared with 0.24 for the route-mean baseline (Supplementary Table~S3).

Collective graph deformation $Q_D$, which has an exact all-scale displacement-wavelet representation, tracks corrected stabilisation ($\rho=0.664$, $n=412$; Fig.~\ref{fig:grounding}c), with within-route correlations of 0.55--0.72. $Q_D$ and maximum atomic displacement give held-out $R^2=0.59$ and 0.60, respectively; ridge regression on 205 structural-response descriptors reaches 0.70 (Supplementary Fig.~S10c). The fingerprints identify the structural transformation, while the electronic and deformation coordinates resolve variation in localisation and stabilisation within each route.

The potential and spin-polarisation fingerprints resolve complementary changes (Supplementary Figs.~S11 and S12). Median relative response is largest for induced K centres in the structural family and for polarons in the spin family, and smallest for native K centres in all three families. Spin-fingerprint change also correlates with localisation gain ($\rho=0.786$, $n=412$).

\subsection{Charge-driven conversion dominates the deepest traps}
\begin{figure}[ht]
\centering
\includegraphics[width=\textwidth]{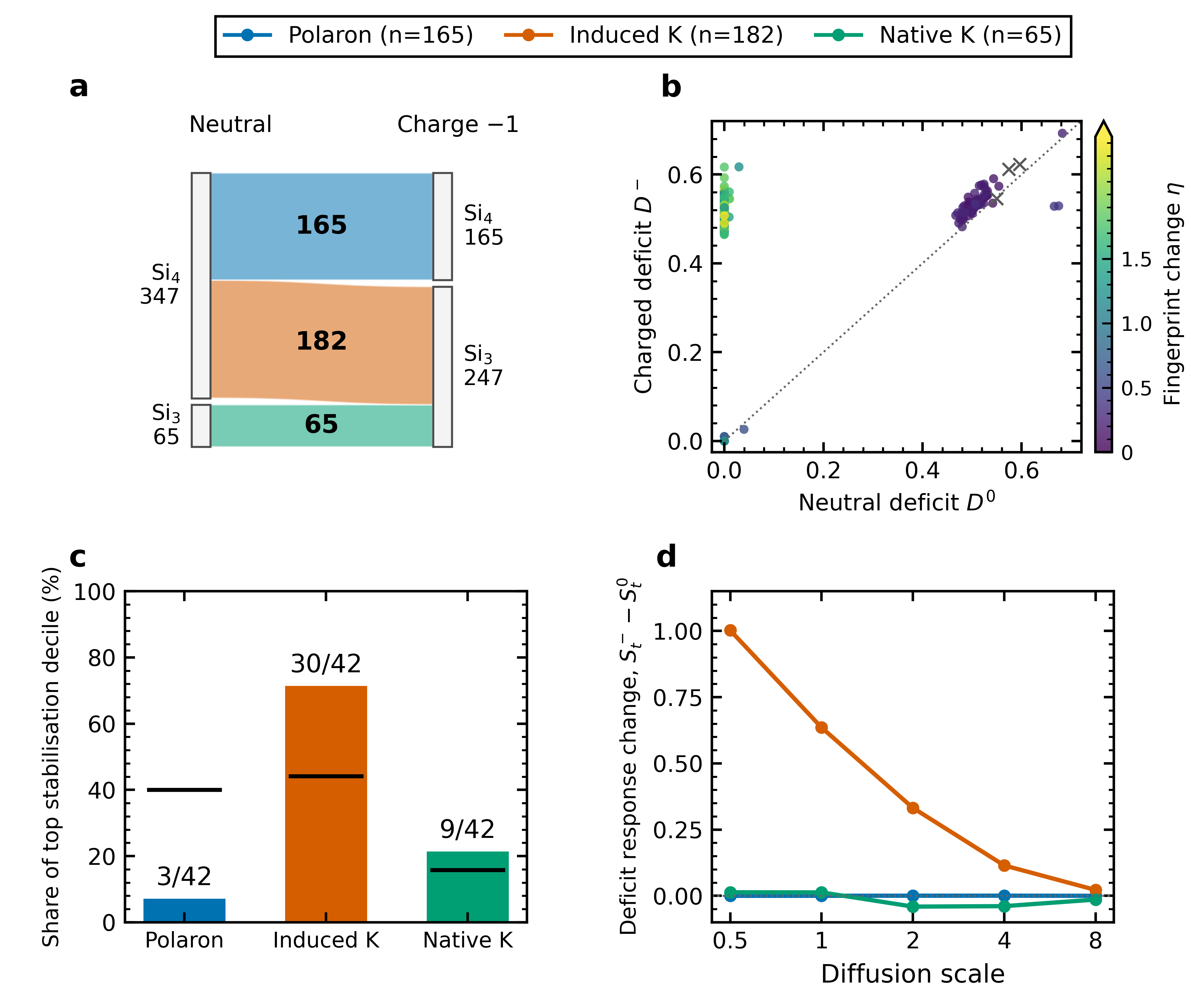}
\caption{\textbf{Chemical interpretation of the graph-wavelet fingerprints in \asirn.} \textbf{a,} Coordination flow from neutral precursor to trapped endpoint. Of 347 Si$_4$ precursors, 182 convert to Si$_3$ and 165 remain Si$_4$; 65 native Si$_3$ sites remain Si$_3$, giving 165 Si$_4$ and 247 Si$_3$ endpoints. The contact classes are defined in Methods. \textbf{b,} Neutral and charged neighbourhood deficits coloured by $\eta$; the three unclassified cells are marked by grey crosses. \textbf{c,} Enrichment of the top decile of corrected stabilisation by chemical route. The black horizontal markers show route fractions in the full classified trap population. \textbf{d,} The induced-K class shows the largest positive median change $S^-_{1,t,{\rm deficit}}-S^0_{1,t,{\rm deficit}}$ of the coordination-deficit scattering response at every scale, completing the link between graph fingerprint and chemistry readout. Bands show 95\% whole-cell bootstrap intervals.}
\label{fig:chemistry}
\end{figure}

Charge-induced Si--N bond breaking converts 182 of 347 Si$_4$ precursors (52.4\%) into Si$_3$ trapped centres (Fig.~\ref{fig:chemistry}a; Supplementary Fig.~S6 and Table~S5). This induced-K class shows the largest relative structural fingerprint change and is strongly enriched in the upper stabilisation tail (Supplementary Fig.~S1 and Table~S2).

The coordination-deficit plane in Fig.~\ref{fig:chemistry}b resolves the change in local chemistry. Polarons cluster near zero deficit in both charge states, native K centres retain a deficit, and induced K centres predominantly gain deficit on charging. Induced K centres account for 30 of the 42 traps in the upper corrected-stabilisation decile (71.4\%), compared with 44.2\% of the 412 classified traps (Fig.~\ref{fig:chemistry}c). Polarons contribute only 3 of the 42 traps (7.1\%), despite making up 40.0\% of that population. Ranking the same 412 traps by corrected spectral trap depth places 29 induced K centres among the 42 deepest traps (69.0\%).

Figure~\ref{fig:chemistry}d links the multiscale response to coordination loss (Supplementary Fig.~S8). The induced-K family shows the largest positive median change of the $S_1$ coordination-deficit channel at every diffusion scale, consistent with creation of an undercoordinated endpoint. Together, the paired electronic, structural and energetic evidence identifies charge-induced Si--N bond breaking as the dominant route to the deepest traps in this ensemble.

\section{Discussion}
State-aligned graph-wavelet fingerprints establish a quantitative, multiscale and multimetric framework for resolving charge-trap formation. Alignment of the trap state before and after localisation connects changes in localisation and electronic-centre position to paired fingerprints of the responding network. These resolve the magnitude, chemical character, scale dependence and spatial extent of network reorganisation, distinguishing the formation and stabilisation of different trapping environments. Each trapping route is thus characterised through the coupled evolution of the electronic state and its atomic surroundings.

In \asirn{}, intrinsic-like, defect-stabilised and conversion-driven trapping coexist within the same amorphous ensemble. The dominant deep-trap route is charge-driven conversion: Si--N bond breaking transforms 182 of 347 Si$_4$ precursors (52.4\%) into undercoordinated trapped centres. This induced-K family accounts for 69.0\% of the deepest spectral-trap decile and 71.4\% of the upper corrected-stabilisation decile, compared with 44.2\% of the classified traps. Charge capture therefore creates the undercoordinated centres that dominate the deepest-trap population in this ensemble. The associated coordination loss and surrounding network reorganisation give this defect-formation process a distinct multiscale signature.

The physical readouts connect these structural transformations to carrier localisation and energetic stabilisation. Occupation-induced level shifts and spin-fingerprint changes track localisation gain, whereas structural fingerprint change tracks local ionic relaxation and graph deformation tracks corrected stabilisation. The structural, potential and spin-polarisation fingerprints retain the distinct physical responses of each precursor-trap pair within a common graph representation. Together with the exact displacement-wavelet representation of $Q_D$, this connects local chemical change, electronic reorganisation and collective ionic relaxation within one quantitative treatment.

Application to other materials follows the same construction. Atomic connectivity and physically defined channel families encode the material's bonding and electronic structure, while the diffusion operators and paired readouts retain their form. A reference normalisation shared by the compared endpoints preserves a consistent measure of change. This provides a common means of identifying precursor environments, resolving their response to charge capture and relating the resulting transformations to localisation and stabilisation. The accompanying implementation and machine-readable data provide a reproducible starting point for adapting this treatment to other disordered materials.

Where hopping rates or kinetic barriers are available, the same diffusion logic can be extended from structural fingerprinting to transport-aware analysis. State-aligned graph-wavelet fingerprints thereby establish a route-resolved structure-property framework for identifying the coupled electronic and atomic transformations that create and stabilise charge traps.

\section{Methods}

\subsection{Amorphous model generation and electronic-structure calculations}
Amorphous \asirn{} starting structures were generated using the \texttt{Vitriflow} melt-quench workflow \cite{COTTOM2026115098}, which uses \textsc{LAMMPS} as the molecular-dynamics engine \cite{Plimpton1995,LAMMPS}. Stoichiometric cells were equilibrated and cooled under NPT conditions using a Nos\'e-Hoover thermostat/barostat \cite{Nose1984,Hoover1985,Martyna1994,Parrinello1981,Shinoda2004} at a cooling rate of 1~K~ps$^{-1}$. The empirical stage employed the Marian-Gastreich two-body (MG2) potential \cite{Marian2000MG2}, selected because comparative benchmarking for amorphous silicon nitride identified MG2 as one of the most reliable starting points for generating low-defect networks prior to DFT refinement \cite{Dasmahapatra2018, Huckmann_22}. Ensemble convergence was assessed from bond-length and bond-angle distributions, pair-correlation functions, ring statistics, density, and internal-void metrics. The resulting neutral cells were then fully relaxed at the hybrid DFT level, allowing both lattice vectors and ionic positions to vary. Charged structures were relaxed at fixed neutral-cell lattice vectors.

All electronic-structure calculations were performed spin-polarised at the $\Gamma$-point with the \textsc{Quickstep} module of \textsc{CP2K} (v2023.1), which implements the mixed Gaussian-and-plane-wave (GPW) formalism \cite{CP2K} \. DZVP-SR-MOLOPT Gaussian basis sets \cite{MOLOPT_Basis_set} and GTH-pseudopotentials \cite{GTH1, GTH2}. Auxiliary plane-wave cutoffs of 650~Ry and 70~Ry were employed. Exchange and correlation were described with the screened hybrid HSE06 functional \cite{HSE06_original,HSE06_erratum}, with the auxiliary density matrix method \cite{ADMM_2010} were used throughout. Geometry and cell optimisations used an energy-convergence threshold of $10^{-7}$~eV and a maximum-force threshold of 0.005~eV\,\AA$^{-1}$.

\subsection{State-resolved neutral/charged dataset construction}
The deposited dataset contains 455 pairs of relaxed neutral and negatively charged cells. Forty pairs were excluded for calculation artefacts or inconsistent electronic-state records: unconverged or non-ground-state correction single points (14 pairs); trap-state Si fractions differing by more than 0.15 between IPR and spectral outputs (15); occupied vertical-anion $\alpha$ levels above empty levels (7); or different maximum-spin atoms in vertical-anion calculations from the relaxation and correction campaigns (9). Five pairs met multiple criteria. The resulting 415-pair dataset retains strain-releasing relaxations and weak-margin traps (Supplementary Table~S1). For each cell, spin-resolved Kohn-Sham frontier states were extracted for both charge states, retaining the common window of states 591--691 per spin (state 641 and 50 states on either side), together with atom-projected quantities, Mulliken populations\cite{Mulliken1955}, and site-level graph descriptors. The neutral $\alpha$-precursor and relaxed trap were matched through their site-resolved orbital character, electron-density distributions and IPR. The relaxed-state comparison covered the 31 highest occupied $\alpha$ orbitals. The correction single points confirmed the state assignments. The atom-projected weights underlying $I_{\rm atom}$ connect each state to the atomic sites over which it is distributed. A paired transition table was then constructed for the aligned trap state before and after localisation, retaining the $(\mathrm{spin},\mathrm{mo\_index})$ label at each endpoint. The neutral $\alpha$/mo641 precursor was paired with the aligned occupied state selected in the relaxed charged cell, preserving state identity when frontier ranks reorder or localised states move strongly in energy.

The local graph response was evaluated at the final trap atom, using the same atom identity in both endpoint structures. Atom-resolved Mulliken data were propagated for chemistry annotation, spin-field descriptors and supplementary audit, but were not used to define the state alignment itself.

Hartree-potential grids were reduced to the potential energy of an electron averaged within 1.0~\AA{} of each nucleus and over the shell 1.4--3.0~\AA{}, excluding shell points within 0.6~\AA{} of any other nucleus. The potential grid was Gaussian-smoothed with a width of 0.15~\AA{}. Site and shell averages used the full grid (0.065~\AA{} spacing). Each average was referenced to the mean for the same species in the same cell. Neutral-state potentials were evaluated at both endpoint geometries. Signed Mulliken spin populations and integrals of the absolute spin density over atomic Voronoi cells were taken from the vertical anion at the neutral geometry and the relaxed anion. Voronoi integrals assigned every grid point to its nearest nucleus. The precursor participation region contains the $\lceil N_{\rm eff}^{s}\rceil$ atoms with largest $|s_i|$, where $s_i$ is the vertical-anion Mulliken spin population and $N_{\rm eff}^{s}=(\sum_i|s_i|)^2/\sum_i s_i^2$. The final trap lies outside this region when its atom is not in that set.

\subsection{Localisation metrics, mobility edges, and relaxation observables}
Electronic localisation was quantified by the atom-projected inverse participation ratio

\[
I_{\rm atom}=\sum_i p_i^2,\qquad \sum_i p_i=1,\qquad
N_{\rm eff}^{\rm el}=I_{\rm atom}^{-1},
\]

where $p_i$ is the normalised atom-projected orbital weight. Localisation gain is $\Delta\log_{10}I=\log_{10}(I^-/I^0)$, using the neutral state and the aligned relaxed trap state. The fixed-nuclei occupation response is $-\delta\epsilon_{\rm occ}=\epsilon^0_{\alpha641}-\epsilon^{-,\rm vert}_{\alpha641}$, with both levels aligned to the neutral N 2s centroid, defined by the mean energy of $\alpha$ states 1--160. The level shift $-\delta\epsilon_{\rm occ}$ was evaluated before and after defect-specific eigenvalue correction. Its correlation with localisation gain is $\rho=0.765$ for the occupation response shown in Fig.~\ref{fig:grounding}a and 0.760 after correction (Supplementary Table~S3). Energy-resolved localisation and elemental-character profiles used 0.1~eV bins, taking a median within each cell and then medians and interquartile ranges across cells. The Si fraction is the summed Si projection divided by the combined Si and N projections. Exemplar densities of states used Gaussian broadening with standard deviation 0.10~eV.

The neutral mobility edges in Fig.~\ref{fig:alignment} are the highest occupied and lowest unoccupied $\alpha$ states with $N_{\rm eff}^{\rm el}\geq56$, one fifth of the cell. The same neutral edges reference both spectra (Supplementary Fig.~S2 and Table~S4). Spectral trap depth is $D_{\rm trap}=\epsilon_{\rm C}^{0}-(\widetilde{\epsilon}_{\rm trap}^{-}+\delta\epsilon_{\rm trap}^{-})$, where $\epsilon_{\rm C}^{0}$ is the neutral conduction mobility edge under the same participation criterion, $\widetilde{\epsilon}_{\rm trap}^{-}$ is the N~2s-aligned relaxed trap level, and $\delta\epsilon_{\rm trap}^{-}$ is its defect-specific eigenvalue correction.

The corrected stabilisation energy was calculated from the vertical and relaxed anion total energies,

\[
E_{\rm stab}^{\rm corr}
=E^-(R^0)+C_{\rm vert}-E^-(R^-)-C_{\rm rel},
\]

where each endpoint correction combines a defect-charge-density image-interaction term and potential alignment. The clamped-ion relative dielectric constant $\varepsilon_\infty=4.2$, describing the purely electronic response. The corresponding static relative dielectric constant, including ionic polarisation, is $\varepsilon_{\rm s}=7.8$. Finite-size corrections were evaluated using a single-shot, density-based potential-correction treatment following Durrant et al.\cite{Durrant2018} and related to the potential-correction formulation of SCPC.\cite{ChagasDaSilva2021} The FNV scheme served as an established control.\cite{Freysoldt2009, Kumagai2014FiniteSize} The corrections were computed with the Olsson Materials Modelling implementation (\url{https://github.com/Olsson-Materials-Modelling}). Total-energy corrections and defect-level eigenvalue shifts were applied separately; the total-energy lowering is distinct from a Kohn-Sham level shift\cite{Naik2018CoFFEE}.

Ionic displacements $u_i$ were obtained by minimum-image differencing followed by subtraction of the whole-cell mean translation. Local relaxation is $u_{\rm rms}=[|\mathcal N|^{-1}\sum_{i\in\mathcal N}|u_i|^2]^{1/2}$ in the fixed neutral 3 \AA{} shell about the trap atom. Displacement participation is $N_{\rm eff}^{u}=(\sum_i|u_i|^2)^2/\sum_i|u_i|^4$, and collective deformation is

\[
Q_D=\sum_{\{i,j\}\in E^0}w_{ij}|u_i-u_j|^2,
\]

using each neutral Si--N edge once. $Q_D$ is an unfitted geometric measure with units \AA{}$^2$; its exact displacement-wavelet representation is given below.

\subsection{Atomistic graph construction and diffusion-wavelet descriptors}
Each endpoint was converted to a periodic Si--N contact graph using a 2.33~\AA{} cutoff and Gaussian weights $w_{ij}=\exp[-(r_{ij}/\sigma)^2]$, with $\sigma=1.25$ \AA{}. The symmetric normalised Laplacian is $L=I-D^{-1/2}AD^{-1/2}$, where $D_{ii}=\sum_jA_{ij}$. The averaging diffusion operator is
\[
H_t=D^{-1/2}e^{-tL}D^{1/2},
\]
whose rows sum to one, so that $H_t\mathbf{1}=\mathbf{1}$ and $(H_t-H_{2t})\mathbf{1}=0$.

The seven atom-level structural channels comprise coordination deficit and excess, mean bond length, bond-length spread, outer-rank distance, next-rank gap and angular deviation. For contact count $z_i$ and expected coordination $k_i=4$ for Si or 3 for N, deficit and excess are $\max(k_i-z_i,0)$ and $\max(z_i-k_i,0)$. Bond means, population standard deviations and angular summaries use the shortest $\min(z_i,k_i)$ contacts, while every contact within the cutoff remains in the graph. Outer-rank distance and next-rank gap are $r_{i(k_i)}$ and $r_{i(k_i+1)}-r_{i(k_i)}$, respectively, with ranks taken over all opposite-species atoms. Angular deviation is the RMS of $\cos\theta-c_i^\ast$ over neighbour pairs, with $c_{\rm Si}^\ast=-1/3$ and $c_{\rm N}^\ast=-1/2$.

The potential family contains the referenced site potential and shell-minus-site contrast. The spin family contains the signed Mulliken spin population and the natural logarithm of the Voronoi integral of absolute spin density, expressed in electron units.

The structural channels were standardised using their means and standard deviations over all 116,200 neutral atoms. Each potential and spin channel was centred on its species mean over the reference-geometry ensemble and divided by its standard deviation over all reference atoms after centring. The same transformations were applied to both endpoints. First-order scattering responses were then evaluated as

\[
S_{1,t}=H_t\left|\left(H_t-H_{2t}\right)X\right|,
\qquad t\in\{0.5,1,2,4,8\},
\]

where $X$ is the standardised channel matrix and the modulus is elementwise. Each band was pooled at its own diffusion scale, and the operators were evaluated from the eigendecomposition of $L$. Its RMS pooling radius at trap atom $i$ is $[\sum_j(H_t)_{ij}r_{ij}^2]^{1/2}$, with median values of 1.24, 1.75, 2.46, 3.45 and 4.73~\AA{} over the 830 endpoint graphs at the five scales. This construction follows diffusion-wavelet and graph-scattering formalisms in a non-kinetic setting \cite{CoifmanMaggioni2006,Hammond2011,Gama2019DiffusionScattering,Gao2019GeometricScattering}.

At the fixed trap atom, the seven structural components at scale $t$ form $v_t^q$ for endpoint $q$. With $R_t^q=\|v_t^q\|_2$ and $V^q$ the concatenation of all five vectors (35 components), the neutral amplitude and relative fingerprint change are

\[
T^0=\sum_tR_t^0,\qquad
\eta=\frac{\|V^--V^0\|_2}{\|V^0\|_2}.
\]

Scale-resolved changes are $\eta_t=\|v_t^--v_t^0\|_2/R_t^0$, and $T^-/T^0$ measures the amplitude change. The same readouts were calculated separately for the two-channel potential and spin families, each with ten components across the five scales; their endpoints denote the reference and relaxed geometries described above. The spatial profile in Fig.~\ref{fig:wavelets}d groups each atom's own $\|V_i^--V_i^0\|_2$ by its unweighted neutral-graph distance from the trap atom, taking shell medians within each cell pair before summarising by route. First-shell response fractions sum the squared structural coefficient changes over the trap atom and its neighbours on the neutral graph, divided by the whole-cell sum at each scale. At the trap atom, channel contributions were calculated for each cell pair from squared coefficient changes summed over scales, normalised by the family total and then averaged within each route.  The raw-centre and raw-shell controls are RMS changes of the seven standardised structural channels at the trap atom and over the fixed neutral 3~\AA{} shell, respectively. The lowpass control is the RMS change of $H_{16}X$ at the trap atom, using each endpoint graph. The unnormalised wavelet control is $\|V^--V^0\|_2/\sqrt{35}$; the geometry-only control retains the five bond and angle channels.

The deformation measure $Q_D$ admits an exact all-scale wavelet identity. Writing $U$ for the displacement matrix and using the neutral $L$ and $D$,

\[
\int_0^\infty
\left\|\left(e^{-tL}-e^{-2tL}\right)D^{1/2}U\right\|_F^2
\frac{{\rm d}t}{t^2}
=(10\ln2-6\ln3)Q_D.
\]

The five-scale displacement-wavelet power in Supplementary Fig.~S10c was calculated as $\sum_{t\in\{0.5,1,2,4,8\}}\|D^{1/2}(H_t-H_{2t})U\|_F^2/t$ on the neutral graph.  The multidescriptor model in Supplementary Fig.~S10c used ridge regression on 215 structural-response descriptors: 108 local displacement, coordination, bond, angle and strain-change descriptors; 90 displacement-wavelet powers, participations and shares; and 17 graph quadratic forms, evaluated over the cell and within 3, 4 and 5~\AA{} of the trap atom. Columns duplicated up to sign were removed within each training fold, leaving 205 of the 215 candidate descriptors; route labels, fingerprint coefficients and electronic quantities were not inputs. Median imputation and standardisation were fitted within training folds. The ridge penalty was selected from 16 logarithmically spaced values by four-fold inner cross-validation within five outer folds, repeated over 20 outer-fold assignments.

\subsection{Chemical routes and coordination readout}
The trap assignments distinguish polarons (Si$_4\!\rightarrow$Si$_4$, $n=165$), induced K centres (Si$_4\!\rightarrow$Si$_3$, $n=182$) and native K centres (Si$_3\!\rightarrow$Si$_3$, $n=65$). These routes distinguish polaronic localisation, charge-induced Si--N bond breaking and capture at a pre-existing Si dangling bond, respectively. Three cells have no assigned route and remain separate. The Si$_4$/Si$_3$ labels denote these chemical motifs; raw cutoff-contact counts remain diagnostic quantities. The continuous neighbourhood deficit in Fig.~\ref{fig:chemistry}b is the Gaussian-weighted mean of $\max(k_i-z_i,0)$ within 3 \AA{} of the fixed trap atom, with weights $\exp[-(r_i/1.5\,\text{\AA})^2]$ and expected coordination $k_i=4$ for Si and 3 for N. It does not assign the routes (Supplementary Fig.~S7). The deficit-channel response is reported as $\Delta S_{1,t,{\rm deficit}}=S^-_{1,t,{\rm deficit}}-S^0_{1,t,{\rm deficit}}$, the change in standardised scattering response at the trap atom.

\subsection{Statistics and software}
Correlations were summarised with Spearman rank coefficients using all valid cell pairs in each declared analysis cohort. Percentile-trend panels use $100\,\mathrm{rank}_{\rm average}(x)/n$, with deciles defined by the ceiling of percentile divided by ten; medians and interquartile ranges were computed before applying display limits. Route-profile confidence intervals use 1,500 whole-cell bootstrap resamples within each route. Stabilisation enrichment compares route fractions in the upper corrected-stabilisation decile with those in the full classified trap population. Pairwise separation by a single readout was summarised by the area under the receiver-operating-characteristic curve (AUC), reported as $\max(\mathrm{AUC},1-\mathrm{AUC})$. Route assignment in the $(T^0,\eta)$ plane used 15-nearest-neighbour classification with five-fold stratified cross-validation, averaged over five random fold assignments. Pairwise vector comparisons used $L_2$-regularised logistic regression ($C=1$) with one fixed five-fold stratified assignment. Classifier inputs were standardised within each training fold. The 95\% classification intervals used 2,000 event bootstrap resamples with held-out scores and fold assignments fixed. The neutral Si--N bond asymmetry at the trap atom served as a reference for separating polarons from induced K centres. Additional controls used the seven unfiltered standardised trap-atom channels with the same scoring procedure, and fingerprint readouts without coordination deficit and excess (Supplementary Table~S7). Whole-cell descriptors in Supplementary Fig.~S9 comprise the mean and standard deviation across atoms of each of the 55 scale-channel coefficients, giving 110 descriptors. For principal-component analysis, their means and standard deviations were fitted to the reference-geometry ensemble and retained for the relaxed endpoint. Each endpoint matrix was then centred separately before decomposition. Within each training fold, the descriptor with the largest absolute Spearman correlation was selected for affine regression and evaluated on the held-out cells. Single-coordinate affine calibrations and route-mean baselines used five outer folds, with coefficients and route means fitted only on the training cells. Their 95\% intervals used 2,000 bootstrap resamples of paired held-out outcomes and predictions, with coordinates and folds fixed. Correlation intervals and paired differences used 2,000 whole-cell resamples shared across readouts within each analysed population. Within-route correlations and held-out calibrations are reported in Supplementary Fig.~S10 and Table~S3. Data processing, statistical analysis and figure generation used custom Python scripts with NumPy, SciPy, pandas, scikit-learn, ASE and Matplotlib \cite{Harris2020NumPy,Virtanen2020SciPy,McKinney2010,ASE2017,Hunter2007}.

\section*{Data availability}
The accompanying machine-readable Supplementary Data contain the paired HSE06 spectra, atomic localisation measures, per-atom structural, potential and spin fields, absolute uncorrected total energies, correction terms, and derived graph-wavelet and physical-response tables. The amorphous model-generation workflow is described in Ref.~\cite{COTTOM2026115098}, and the relaxed structures are freely available at Zenodo [10.5281/zenodo.21111822].

\section*{Code availability}
The analysis and plotting code accompanies the manuscript. The development repository is maintained on GitHub [\url{https://github.com/Olsson-Materials-Modelling}].

\section*{Acknowledgements}
This work was conducted at the Advanced Research Center for Nanolithography, a public-private partnership between the University of Amsterdam (UvA), Vrije Universiteit Amsterdam (VU), Rijksuniversiteit Groningen (RUG), the Netherlands Organisation for Scientific Research (NWO), and the semiconductor equipment manufacturer ASML. This work made use of the Dutch national e-infrastructure with the support of the SURF Cooperative using grant no. EINF-8981, and EINF-20460 and the authors thank SURF (www.surf.nl) for the support in using the National Supercomputer Snellius. This work made use of the ARCNL Minerva cluster hosted at HOPSTER at UvA. E.O. is grateful for a WISE Fellowship from the NWO and acknowledges support via Holland High Tech through a public–private partnership grant in research and development within the Dutch top sector of High-Tech Systems and Materials (HTSM). 

\section*{Author contributions}
E.O. and J.C. conceived the study and developed the methodological framework. J.C. implemented the software and computational workflow, performed the investigation, curated the data, carried out the formal analysis, and produced the visualisations. E.O. supervised the project, administered the research programme, secured funding, and led the overall scientific direction. J.C. prepared the original manuscript draft. E.O. and J.C. validated and interpreted the results, revised and edited the manuscript, and approved the final version.

\section*{Competing interests}
The authors declare no competing interests.

\bibliography{ref}

\end{document}


\maketitle
\section*{Supplementary Figures}

\begin{figure}[p]
\centering
\includegraphics[width=\textwidth,height=0.90\textheight,keepaspectratio]{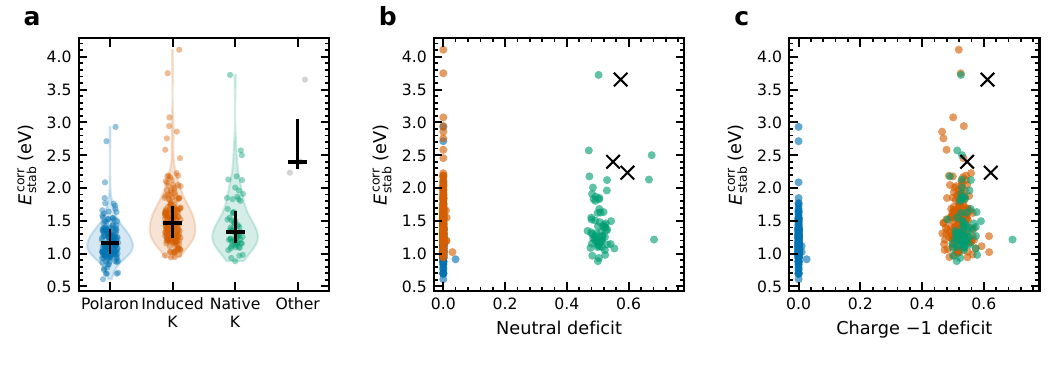}
\caption{\textbf{Corrected stabilisation and local coordination deficit.} \textbf{a,} Corrected stabilisation energies for 415 pairs; points are events, shaded distributions show their spread, and black marks denote medians and interquartile ranges. \textbf{b,c,} Stabilisation against the neutral (b) and relaxed (c) neighbourhood deficit. The pooled charged-deficit correlation ($\rho=0.38$) is dominated by differences between routes; within-route coefficients are $-0.12$, $0.00$ and $0.02$ for polarons, induced K and native K, respectively. Blue, vermillion and bluish-green denote these routes; crosses in b,c mark the three unclassified events. Populations and route summaries are listed in Tables~\ref{tab:si_bookkeeping} and \ref{tab:si_route_summary}.}
\label{fig:si_1}
\end{figure}
\clearpage

\begin{figure}[p]
\centering
\includegraphics[width=\textwidth,height=0.90\textheight,keepaspectratio]{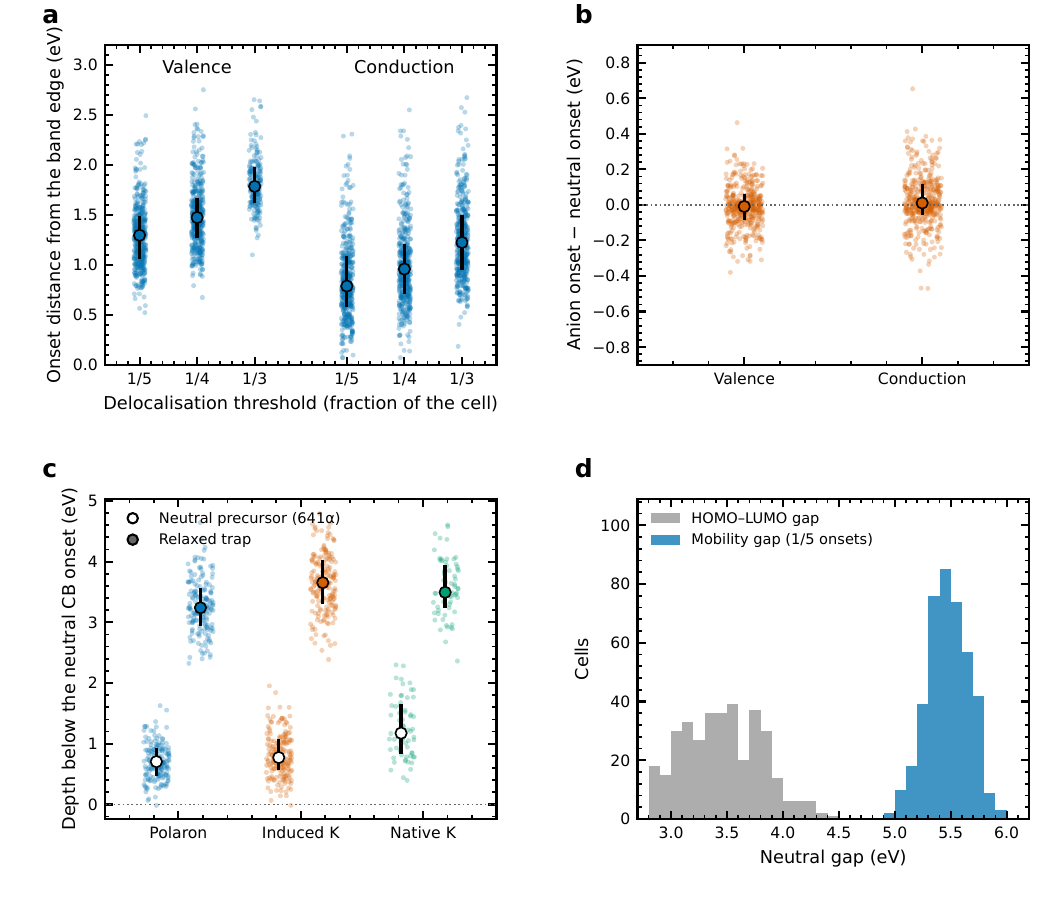}
\caption{\textbf{Mobility-onset sensitivity and spectral trap depths.} \textbf{a,} Neutral valence and conduction onset distances for participation thresholds of one fifth, one quarter and one third of the cell. \textbf{b,} Anion-minus-neutral onset shifts at the one-fifth threshold ($n=412$). \textbf{c,} Precursor and relaxed-trap depths below the neutral conduction onset, shown by open and filled markers, respectively. \textbf{d,} Neutral HOMO--LUMO and mobility gaps. Points are cells; large markers and bars show medians and interquartile ranges. Spectral quantities use the N~2s-aligned scale without defect-specific eigenvalue shifts. Threshold coverage and numerical summaries are given in Table~\ref{tab:si_onsets}.}
\label{fig:si_2}
\end{figure}
\clearpage

\begin{figure}[p]
\centering
\includegraphics[width=\textwidth,height=0.90\textheight,keepaspectratio]{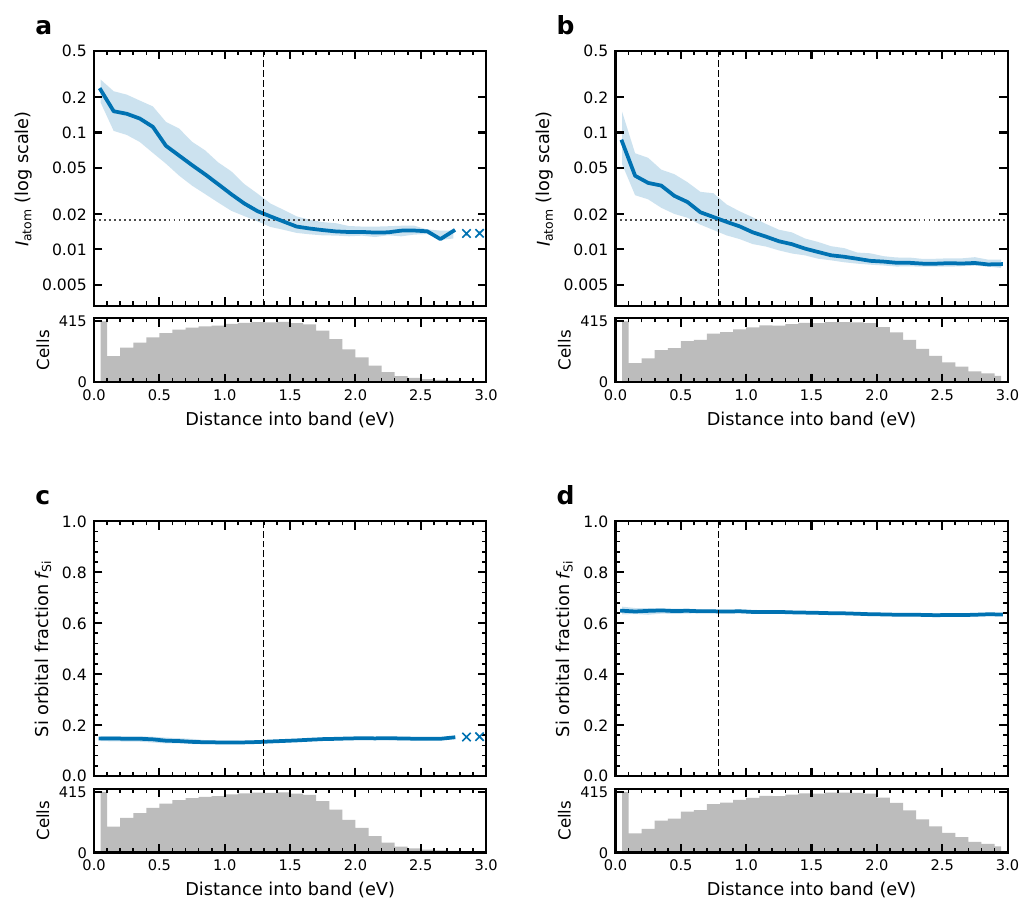}
\caption{\textbf{Neutral localisation and elemental character.} \textbf{a,b,} Atomic IPR of neutral $\alpha$ states against energy distance from the valence and conduction frontiers. \textbf{c,d,} Si-projected fraction of the same states. Lines and bands show medians and interquartile ranges across cell-level bin medians, with equal cell weight. Dotted horizontal lines mark $I_{\rm atom}=5/280$; dashed vertical lines mark the median one-fifth onset distances. Lower strips show the number of contributing cells; crosses identify bins with fewer than ten. Valence states are N-dominated and conduction states are Si-dominated.}
\label{fig:si_3}
\end{figure}
\clearpage

\begin{figure}[p]
\centering
\includegraphics[width=\textwidth,height=0.65\textheight,keepaspectratio]{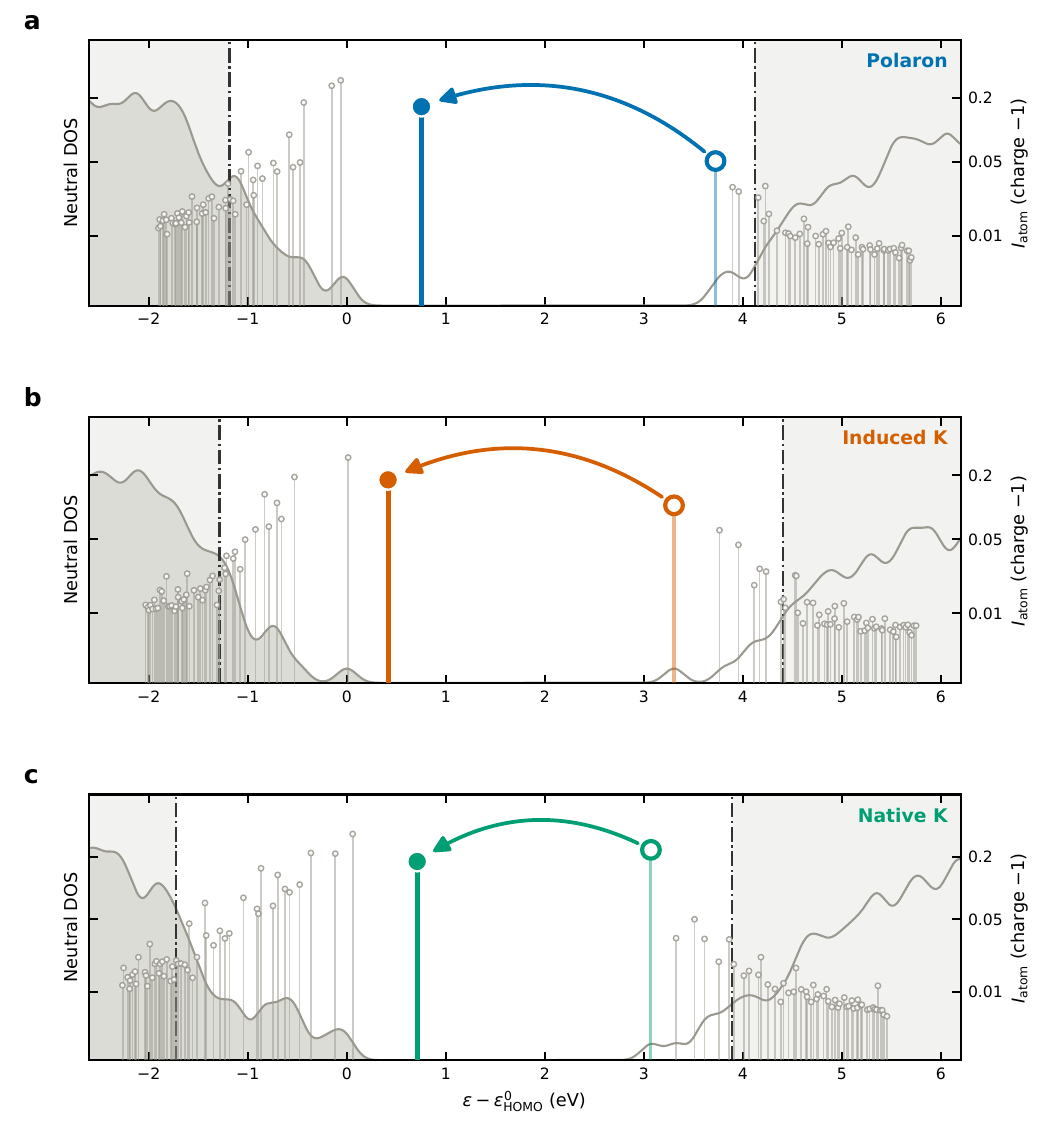}
\caption{\textbf{Spectra and localisation for representative trapping routes.} \textbf{a,} Polaron \texttt{box\_208}. \textbf{b,} Induced K \texttt{box\_030}. \textbf{c,} Native K \texttt{box\_355}. Grey profiles show the neutral $\alpha$ density of states, with occupied states filled; sticks show the relaxed-anion atomic IPR for states 591--691. Open and filled circles mark the neutral precursor and relaxed trap, linked by arrows. Dash-dotted lines and shading identify the neutral mobility edges. Energies are N~2s-aligned and referenced to the neutral $\alpha$ HOMO, without defect-specific eigenvalue shifts. The polaron and induced-K precursors become more localised on trapping; the native-K precursor is already localised.}
\label{fig:si_4}
\end{figure}
\clearpage

\begin{figure}[p]
\centering
\includegraphics[width=\textwidth,height=0.90\textheight,keepaspectratio]{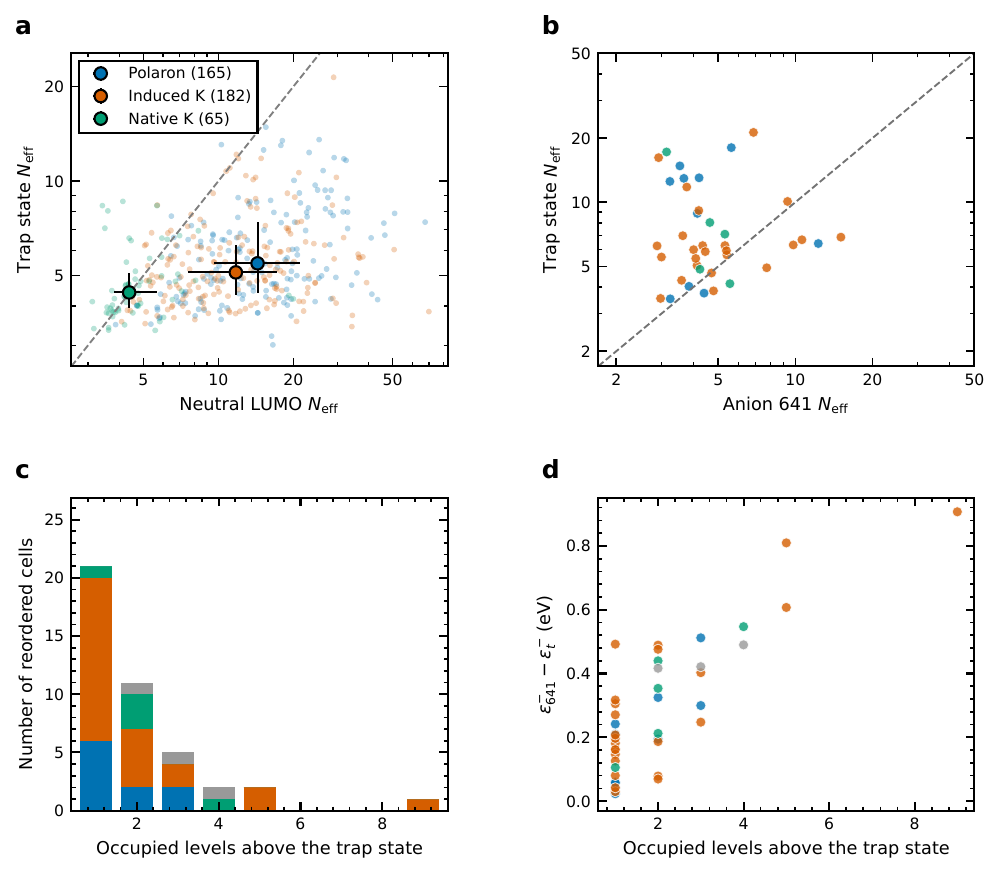}
\caption{\textbf{Localisation and spectral reordering of the aligned trap state.} \textbf{a,} Effective participation of the neutral LUMO ($641\alpha$) and relaxed trap for 412 pairs. Large markers and bars show route medians and interquartile ranges; the diagonal denotes equal participation. \textbf{b,} Participation of anion state 641 against the selected trap for 39 reordered cases with IPR; the 373 unchanged complete pairs coincide on the diagonal and are omitted. \textbf{c,} Occupied states above the trap in the 42 reordered spectra. \textbf{d,} Energy error from substituting anion state 641 for the trap. State indices are one-based.}
\label{fig:si_5}
\end{figure}
\clearpage

\begin{figure}[p]
\centering
\includegraphics[width=\textwidth,height=0.90\textheight,keepaspectratio]{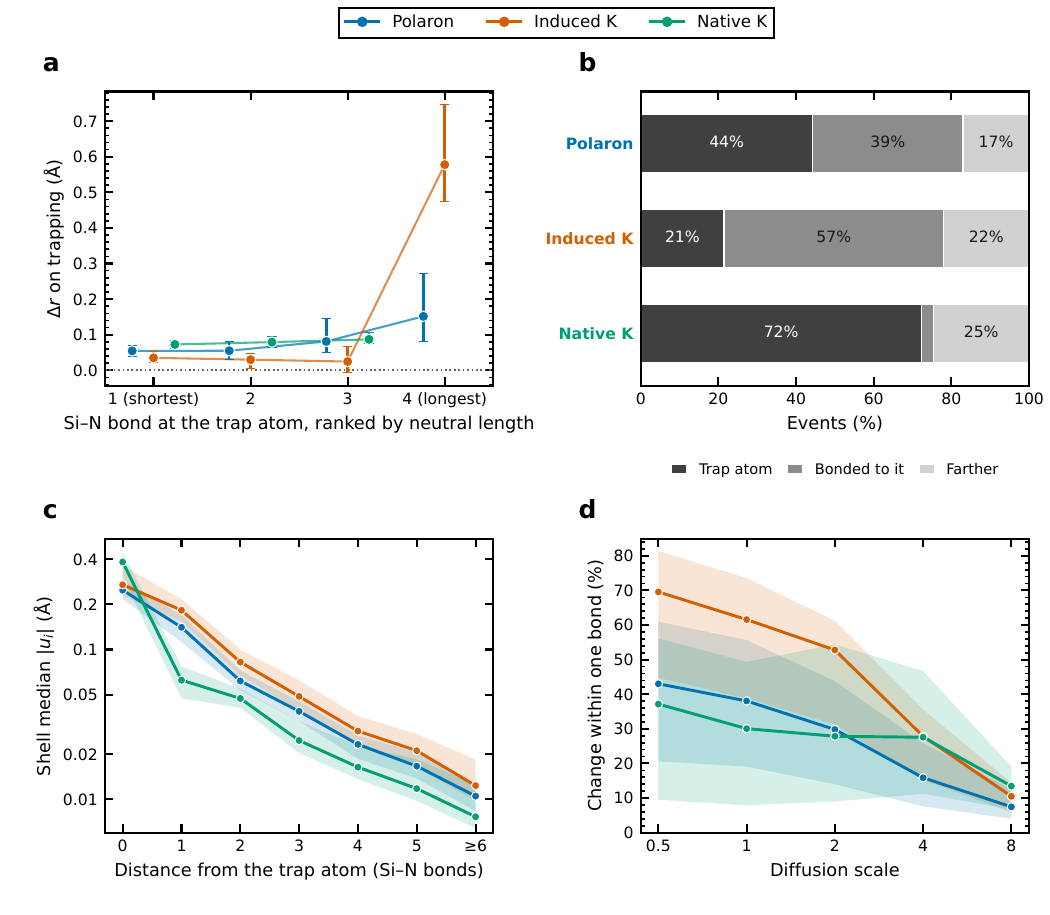}
\caption{\textbf{Bond-resolved and spatial ionic relaxation.} \textbf{a,} Si--N bond-length changes at the trap atom, ranked by neutral length; markers and bars show route medians and interquartile ranges. The longest neutral bond has the largest elongation in 98\% of induced-K events. \textbf{b,} Location of the most displaced atom relative to the trap site. \textbf{c,} Atomic displacement against neutral-graph distance; lines and bands show medians and interquartile ranges of event-level shell medians. \textbf{d,} Share of the squared fingerprint change on the trap atom and its bonded neighbours, with route medians and interquartile bands. All panels use the 412 main-route events; bond statistics are listed in Table~\ref{tab:si_bonds}.}
\label{fig:si_6}
\end{figure}
\clearpage

\begin{figure}[p]
\centering
\includegraphics[width=\textwidth,height=0.90\textheight,keepaspectratio]{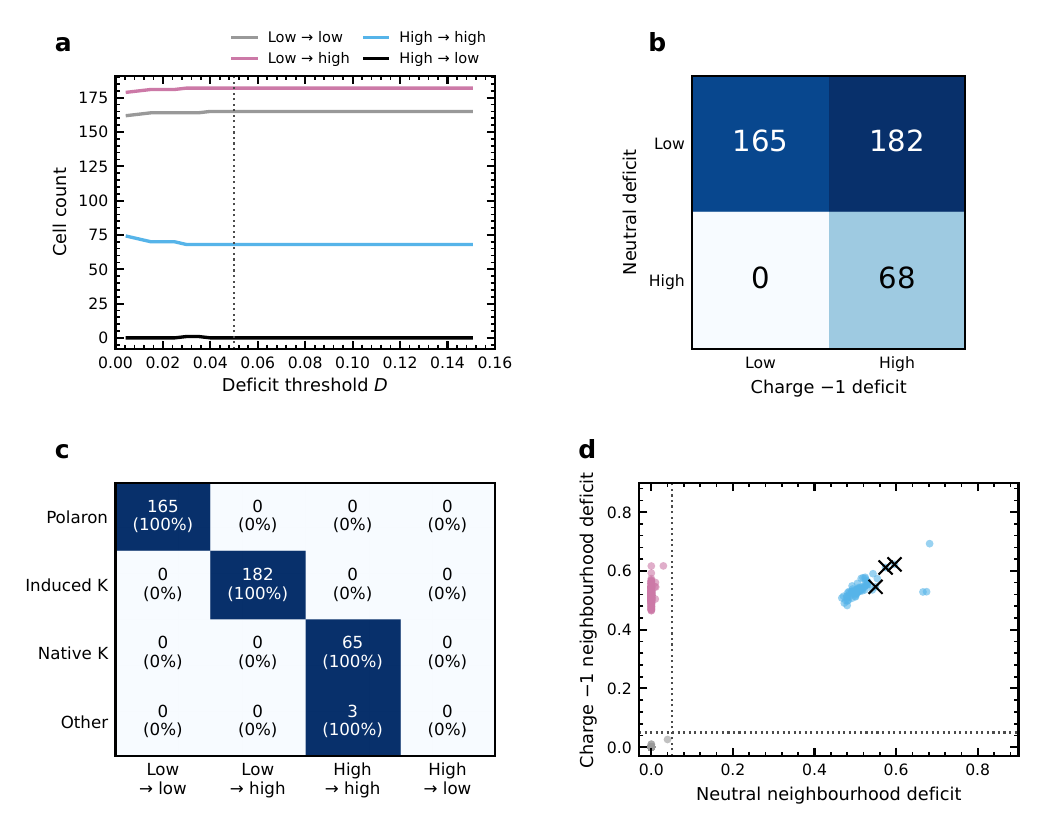}
\caption{\textbf{Sensitivity of the neighbourhood-deficit classification.} \textbf{a,} Diagnostic class populations across deficit thresholds from 0.005 to 0.15. \textbf{b,} Neutral-to-relaxed class counts at $D=0.05$. \textbf{c,} Correspondence between deficit classes and chemical routes. \textbf{d,} Paired neighbourhood deficits, with dotted lines at $D=0.05$ and crosses for the three N-centred events. The comparison includes all 415 analysed pairs. Chemical route labels remain fixed throughout the threshold comparison.}
\label{fig:si_7}
\end{figure}
\clearpage

\begin{figure}[p]
\centering
\includegraphics[width=\textwidth,height=0.90\textheight,keepaspectratio]{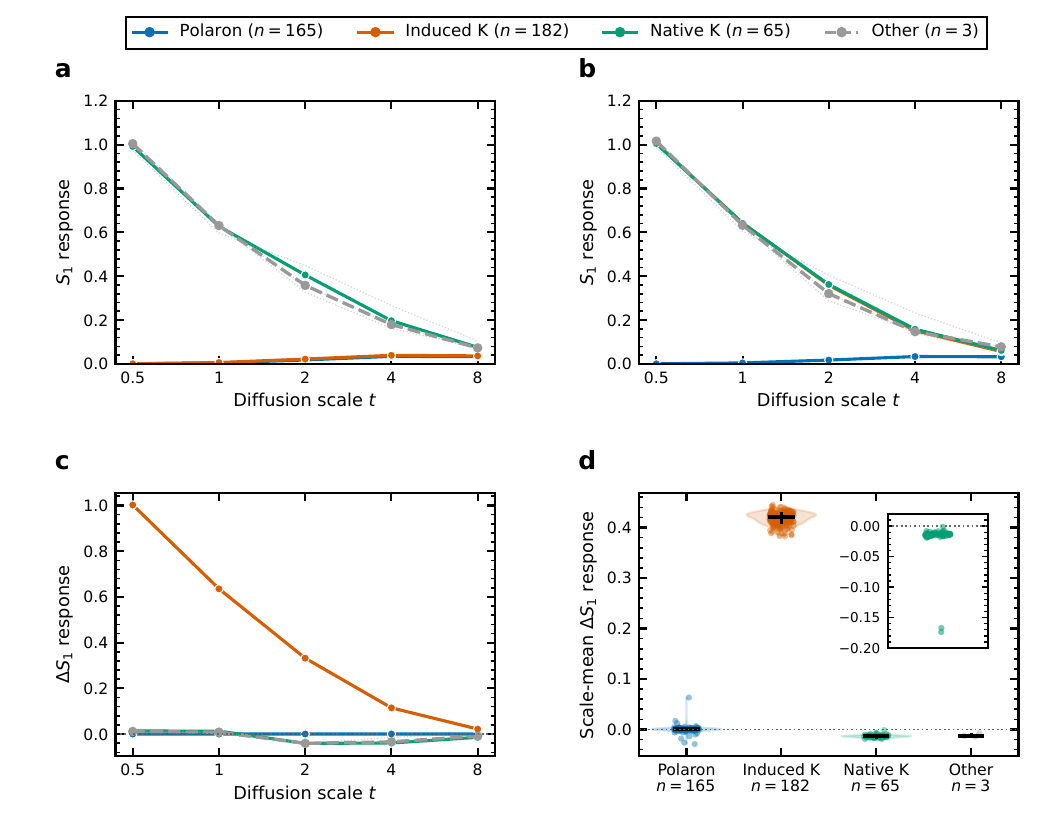}
\caption{\textbf{Coordination-deficit response across graph scales.} \textbf{a,b,} Neutral and relaxed deficit-channel scattering responses. \textbf{c,} Paired response change. Lines and bands in a--c show route medians and 95\% whole-cell bootstrap intervals; dotted traces show the three unclassified events and the dashed line their median. \textbf{d,} Distribution of each event's mean response change across the five scales; black marks show medians and interquartile ranges. Inset: all native-K values, including two events below the main axis in which a two-coordinate N bonded to the trap atom regains its third bond on capture. Induced K centres show the largest positive median change at every scale, while polaron deficit responses remain nearly unchanged.}
\label{fig:si_8}
\end{figure}
\clearpage

\begin{figure}[p]
\centering
\includegraphics[width=\textwidth,height=0.90\textheight,keepaspectratio]{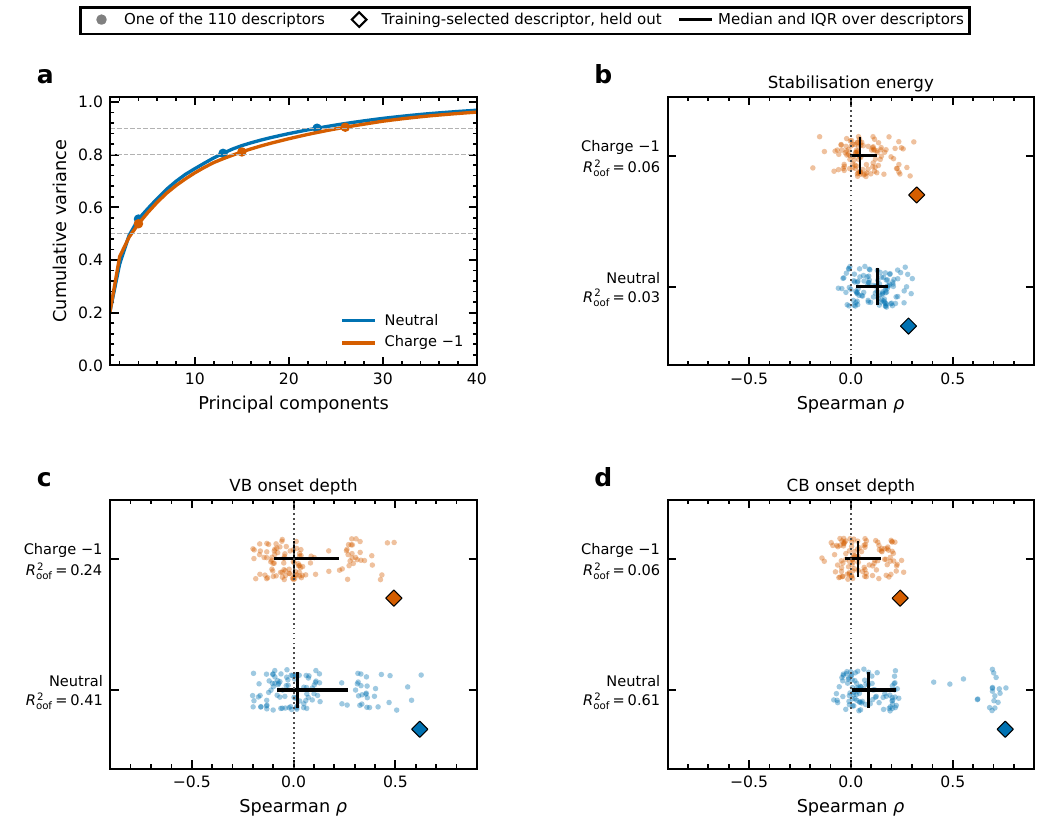}
\caption{\textbf{Whole-graph descriptor variation and physical associations.} \textbf{a,} Cumulative principal-component variance for the 110 whole-graph descriptors at the reference and relaxed geometries ($n=415$). \textbf{b--d,} Descriptor correlations with corrected stabilisation, valence onset depth and conduction onset depth ($n=412$ each). Coloured points show individual Spearman coefficients; black marks give their median and interquartile range across descriptors. Diamonds show correlations between outcomes and held-out single-descriptor predictions; row labels give held-out $R^2$.}
\label{fig:si_9}
\end{figure}
\clearpage

\begin{figure}[p]
\centering
\includegraphics[width=\textwidth,height=0.90\textheight,keepaspectratio]{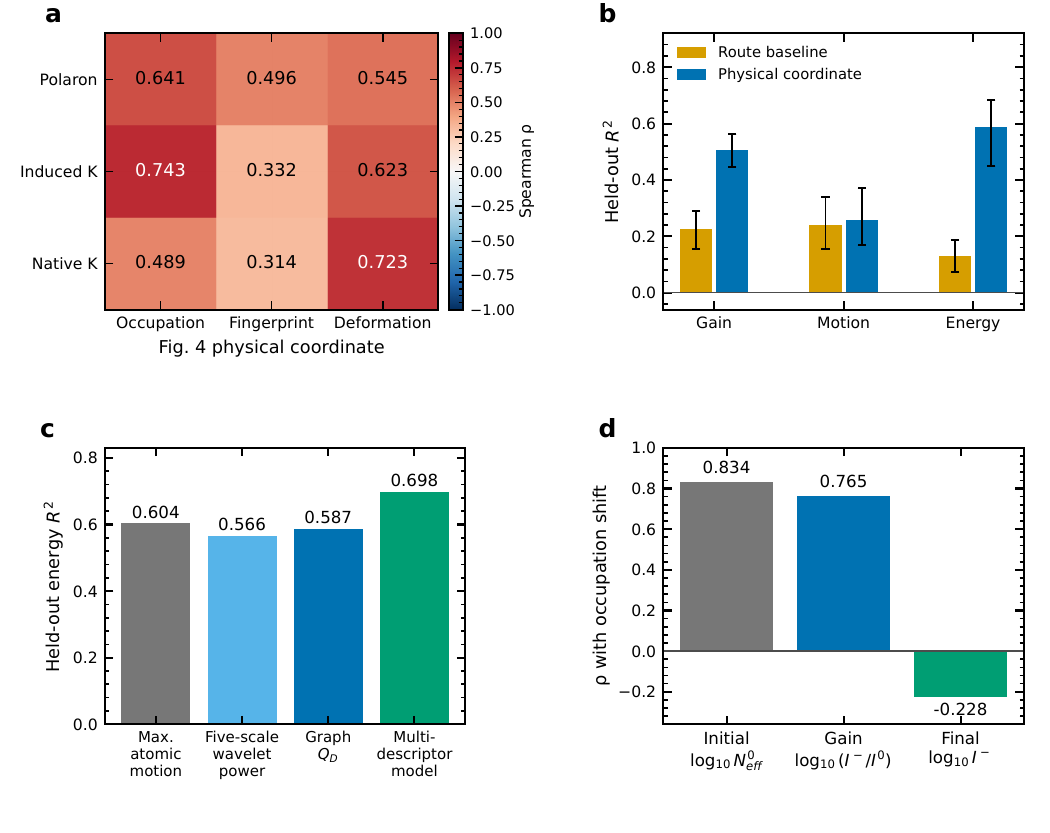}
\caption{\textbf{Physical-coordinate correlations and held-out calibrations.} \textbf{a,} Within-route correlations for the electronic, structural and deformation coordinates of main Fig.~4. \textbf{b,} Held-out $R^2$ for single-coordinate calibrations and route-mean baselines for localisation gain, ionic motion and stabilisation ($n=412$ each). Bars show 95\% paired-cell bootstrap intervals with coordinates and folds fixed. \textbf{c,} Stabilisation calibrations using maximum displacement, five-scale degree-weighted displacement-wavelet power and $Q_D$, alongside a multidescriptor structural model (ridge regression on 205 distinct structural-response descriptors; Methods), all with $n=412$. \textbf{d,} Occupation-shift correlations with initial participation, localisation gain and final IPR on the same 412 pairs. Numerical summaries are listed in Table~\ref{tab:si_coordinates}.}
\label{fig:si_10}
\end{figure}
\clearpage

\begin{figure}[p]
\centering
\includegraphics[width=\textwidth,height=0.50\textheight,keepaspectratio]{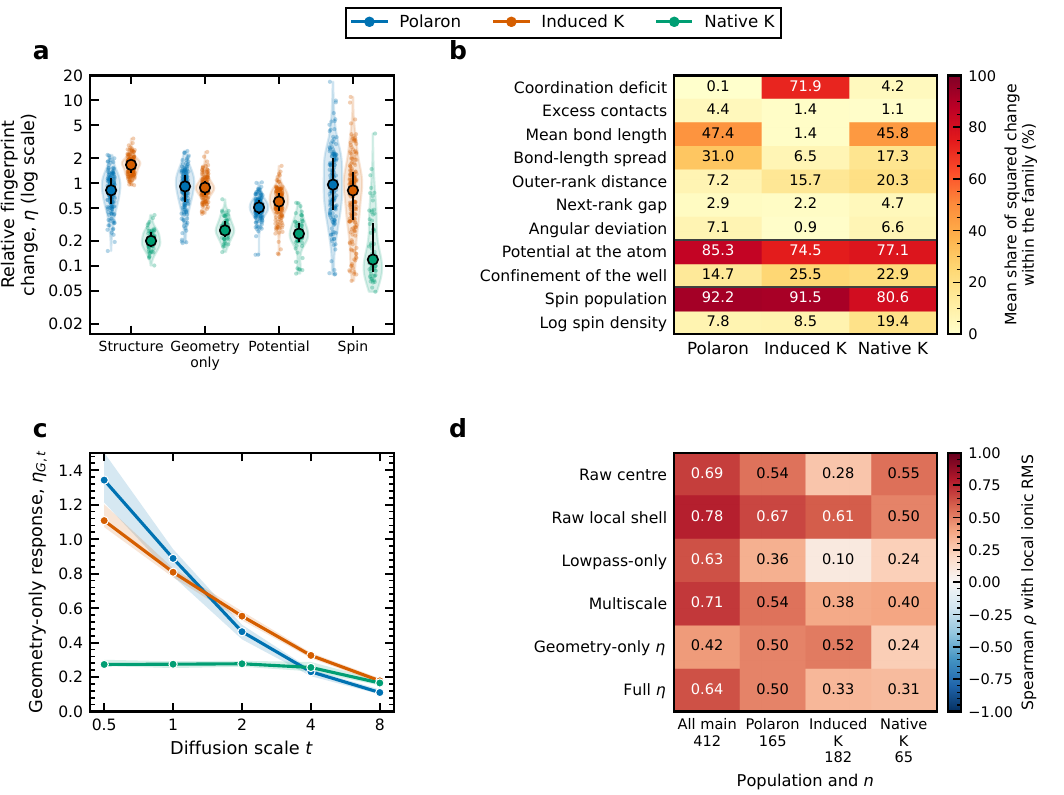}
\caption{\textbf{Channel contributions and structural-response comparisons.} All panels use the 412 main-route events. \textbf{a,} Relative response of the structural, geometry-only, potential and spin families; points are events, with route medians and interquartile bars. Geometry-only retains the five bond and angle channels. \textbf{b,} Mean event-wise share of squared response change assigned to each channel within its family. \textbf{c,} Geometry-only response across scale, with route medians and 95\% whole-cell bootstrap intervals. \textbf{d,} Local ionic RMS correlations for raw centre and local-shell structural changes, lowpass and unnormalised wavelet changes, and geometry-only and full structural $\eta$. Correlation intervals and paired differences are supplied in \texttt{SI\_tables/FigS11d\_correlations.csv}.}
\label{fig:si_11}
\end{figure}
\clearpage

\begin{figure}[p]
\centering
\includegraphics[width=\textwidth,height=0.90\textheight,keepaspectratio]{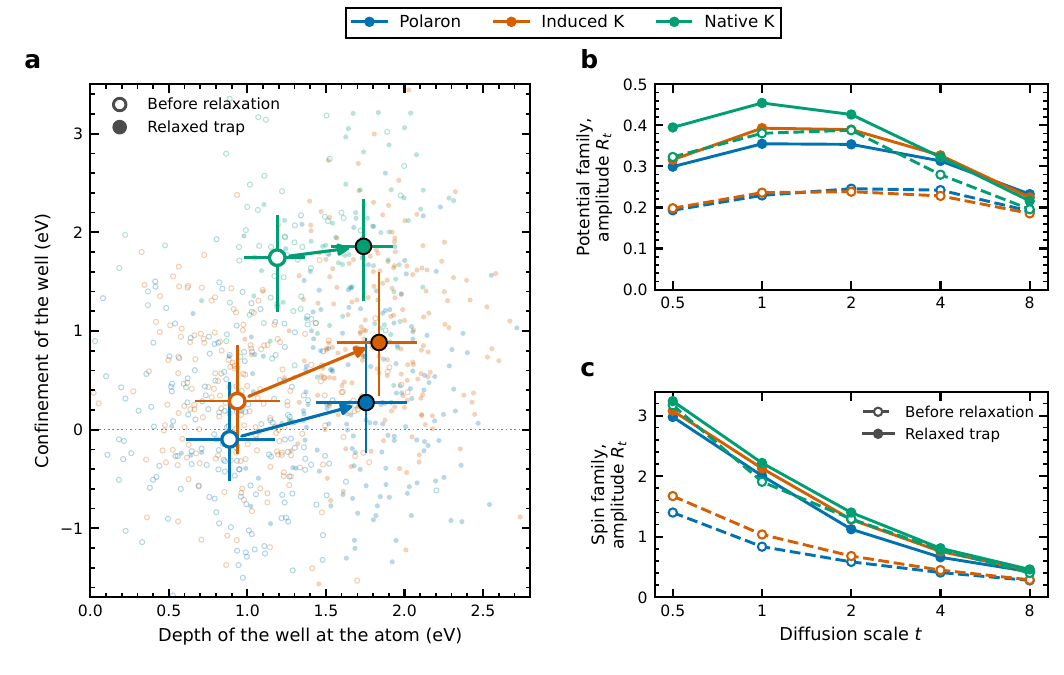}
\caption{\textbf{Potential and spin response at the trapping site.} \textbf{a,} Negative referenced site potential against shell-minus-site potential contrast for the 412 main-route events. Open and filled markers denote reference and relaxed geometries; small points are events, large markers and bars give route medians and interquartile ranges, and arrows connect paired route medians. All events enter the summaries, including points outside the display window. \textbf{b,c,} Route-median scattering amplitudes $R_t$ for the potential and spin families; open dashed curves denote the reference geometry and filled solid curves the relaxed geometry. Potentials are from neutral calculations at both geometries; spin fields are from the vertical and relaxed anions. The relaxed site-potential medians are similar, while the potential contrasts remain route-dependent.}
\label{fig:si_12}
\end{figure}
\clearpage
\section*{Supplementary Tables}

{\small
\setlength{\tabcolsep}{4pt}
\renewcommand{\arraystretch}{1.12}
\begin{longtable}{@{}p{0.34\linewidth}p{0.14\linewidth}p{0.48\linewidth}@{}}
\caption{Analysis populations and state correspondence. Pairs are set aside only where a calculation is flawed or two calculations disagree (Methods); all other entries refer to the analysed pairs.}\label{tab:si_bookkeeping}\\
\toprule
Quantity & Value & Note \\
\midrule
\endfirsthead
\caption[]{Analysis populations and state correspondence. (continued)}\\
\toprule
Quantity & Value & Note \\
\midrule
\endhead
\midrule
\multicolumn{3}{r}{\emph{Continued on next page}}\\
\endfoot
\bottomrule
\endlastfoot
Deposited neutral / charge $-1$ pairs & 455 & Si$_{120}$N$_{160}$, HSE06, 280 atoms \\
Set aside for calculation inconsistencies & 40 & correction single point not converged or not the ground state 14; IPR of another state 15; vertical occupation not aufbau 7; vertical runs disagree on the site 9; four pairs meet more than one criterion \\
Analysed pairs & 415 & used in every figure and table \\
Main-route events & 412 & polaron 165, induced K 182, native K 65 \\
Unclassified cells & 3 & \texttt{box\_006}, \texttt{box\_081\_r}, \texttt{box\_170}: N-anchored; Table~\ref{tab:si_unclassified} \\
Neutral endpoint motifs & 347 / 65 & Si$_4$ / Si$_3$ at the trap atom (main routes) \\
Charge $-1$ endpoint motifs & 165 / 247 & Si$_4$ / Si$_3$ at the trap atom (main routes) \\
Pairs with relaxed-state IPR & 412 & the three unclassified cells lack relaxed-state IPR \\
Trap outside precursor participation region & 7 & 4 polarons and 3 unclassified events; participation-region criterion defined in Methods \\
Trap state not index 641 & 42 & level ordering changed on relaxation \\
Trap below the neutral HOMO & 44 & after N~2s alignment \\
Trap Si-fraction margin below 0.10 & 9 & over the next occupied state among states 611--641; below 0.05 in 3 cells; median margin 0.43 \\
Neutral lower at the anion geometry & 6 & all in the upper corrected-stabilisation decile; without them induced K holds 29 of 41 \\
\end{longtable}
}

\clearpage
{\small
\setlength{\tabcolsep}{4pt}
\renewcommand{\arraystretch}{1.12}
\begin{landscape}
\setlength{\LTcapwidth}{\linewidth}
\begin{longtable}{@{}p{0.40\linewidth}p{0.18\linewidth}p{0.18\linewidth}p{0.18\linewidth}@{}}
\caption{Route-resolved structural, electronic and energetic summaries. Values are medians [interquartile ranges]. Spectral summaries use N~2s-aligned energies without defect-specific eigenvalue shifts. The two separately labelled deciles include the corresponding corrections and each contain 42 of the 412 main-route events.}\label{tab:si_route_summary}\\
\toprule
Quantity & Polaron ($n=165$) & Induced K ($n=182$) & Native K ($n=65$) \\
\midrule
\endfirsthead
\caption[]{Route-resolved structural, electronic and energetic summaries. (continued)}\\
\toprule
Quantity & Polaron ($n=165$) & Induced K ($n=182$) & Native K ($n=65$) \\
\midrule
\endhead
\midrule
\multicolumn{4}{r}{\emph{Continued on next page}}\\
\endfoot
\bottomrule
\endlastfoot
Neutral amplitude $T^0$ & 1.51 [1.30, 1.78] & 1.81 [1.59, 2.17] & 3.48 [3.28, 3.65] \\
Fingerprint change $\eta$ & 0.821 [0.582, 1.125] & 1.665 [1.381, 1.884] & 0.201 [0.173, 0.248] \\
Amplitude ratio $T^-/T^0$ & 1.491 [1.277, 1.729] & 1.696 [1.466, 1.914] & 0.942 [0.915, 0.978] \\
Occupation shift $-\delta\epsilon_{\rm occ}$ (eV) & 0.042 [0.010, 0.073] & 0.023 [-0.035, 0.058] & -0.107 [-0.173, -0.034] \\
Localisation gain $\log_{10}(I^-/I^0)$ & 0.397 [0.219, 0.556] & 0.301 [0.138, 0.498] & -0.002 [-0.041, 0.070] \\
Precursor $N_{\rm eff}^{\rm el}$ (atoms) & 14.4 [9.6, 21.2] & 11.7 [7.6, 17.2] & 4.4 [3.8, 5.7] \\
Trap $N_{\rm eff}^{\rm el}$ (atoms) & 5.5 [4.4, 7.4] & 5.1 [4.3, 6.3] & 4.4 [3.9, 5.1] \\
Local ionic RMS (\AA) & 0.154 [0.134, 0.174] & 0.222 [0.190, 0.256] & 0.151 [0.129, 0.184] \\
Maximum atomic displacement (\AA) & 0.30 [0.25, 0.38] & 0.51 [0.39, 0.67] & 0.40 [0.32, 0.49] \\
Ionic participation $N_{\rm eff}^{u}$ (atoms) & 12.0 [9.2, 15.0] & 9.6 [5.1, 12.3] & 3.8 [2.5, 8.1] \\
Graph deformation $Q_D$ (\AA$^2$) & 0.122 [0.099, 0.188] & 0.240 [0.165, 0.409] & 0.125 [0.098, 0.224] \\
$E_{\rm stab}^{\rm corr}$ (eV) & 1.17 [1.02, 1.35] & 1.47 [1.25, 1.70] & 1.33 [1.18, 1.62] \\
Relaxed trap level, $\epsilon_t-\epsilon^0_{\rm HOMO}$ (eV) & +0.89 [+0.56, +1.28] & +0.55 [+0.18, +0.83] & +0.59 [+0.33, +0.86] \\
Precursor$\rightarrow$trap level drop (eV) & 2.55 [2.28, 2.77] & 2.82 [2.54, 3.06] & 2.26 [1.97, 2.50] \\
Precursor depth below conduction onset (eV) & 0.71 [0.50, 0.91] & 0.77 [0.59, 1.05] & 1.17 [0.87, 1.63] \\
Relaxed depth below conduction onset (eV) & 3.24 [2.97, 3.53] & 3.65 [3.33, 3.99] & 3.49 [3.26, 3.92] \\
Trap outside precursor participation region & 4/165 & 0/182 & 0/65 \\
Median level-drop/$E_{\rm stab}^{\rm corr}$ ratio & 2.2 & 1.9 & 1.6 \\
Level drop--stabilisation Spearman $\rho$ & 0.72 & 0.55 & 0.85 \\
\addlinespace
Deepest corrected spectral decile & 7/42 & 29/42 & 6/42 \\
Upper corrected-stabilisation decile & 3/42 & 30/42 & 9/42 \\
\end{longtable}
\end{landscape}
}

\clearpage
{\small
\setlength{\tabcolsep}{4pt}
\renewcommand{\arraystretch}{1.12}
\begin{longtable}{@{}>{\raggedright\arraybackslash}p{0.245\linewidth}rp{0.06\linewidth}p{0.19\linewidth}p{0.175\linewidth}p{0.175\linewidth}@{}}
\caption{Physical-coordinate correlations and held-out performance. Within-route coefficients are ordered polaron, induced K, native K. Brackets give 95\% paired-cell bootstrap intervals with coordinates and folds fixed. The held-out score is for the single-coordinate affine calibration; the baseline uses route means. The occupation shift $-\delta\epsilon_{\rm occ}$ is N~2s-aligned without a finite-size correction of the occupied level; with that correction the route medians are $-0.50$, $-0.53$ and $-0.68$~eV and the correlation with the gain is 0.760 against 0.765 on the same 412 events.}\label{tab:si_coordinates}\\
\toprule
Coordinate $\rightarrow$ response & $n$ & $\rho$ & $\rho$ within route & Calibration $R^2$ & Route-baseline $R^2$ \\
\midrule
\endfirsthead
\caption[]{Physical-coordinate correlations and held-out performance. (continued)}\\
\toprule
Coordinate $\rightarrow$ response & $n$ & $\rho$ & $\rho$ within route & Calibration $R^2$ & Route-baseline $R^2$ \\
\midrule
\endhead
\midrule
\multicolumn{6}{r}{\emph{Continued on next page}}\\
\endfoot
\bottomrule
\endlastfoot
Occupation shift $\rightarrow$ IPR gain & 412 & 0.765 & 0.64 / 0.74 / 0.49 & 0.51 [0.45, 0.56] & 0.23 [0.16, 0.29] \\
Fingerprint change $\eta$ $\rightarrow$ local RMS & 412 & 0.636 & 0.50 / 0.33 / 0.31 & 0.26 [0.17, 0.37] & 0.24 [0.16, 0.34] \\
Graph deformation $Q_D$ $\rightarrow$ $E_{\rm stab}^{\rm corr}$ & 412 & 0.664 & 0.55 / 0.62 / 0.72 & 0.59 [0.45, 0.68] & 0.13 [0.07, 0.19] \\
\end{longtable}
}

\clearpage
{\small
\setlength{\tabcolsep}{4pt}
\renewcommand{\arraystretch}{1.12}
\begin{longtable}{@{}llrl@{}}
\caption{Neutral mobility-onset sensitivity. Onset distances are measured from the neutral HOMO or LUMO and reported as medians [interquartile ranges]. Counts give the cells with an onset within the analysed window, out of 415.}\label{tab:si_onsets}\\
\toprule
Threshold & Band & Cells & Onset distance (eV) \\
\midrule
\endfirsthead
\caption[]{Neutral mobility-onset sensitivity. (continued)}\\
\toprule
Threshold & Band & Cells & Onset distance (eV) \\
\midrule
\endhead
\midrule
\multicolumn{4}{r}{\emph{Continued on next page}}\\
\endfoot
\bottomrule
\endlastfoot
1/5 & Valence & 415 & 1.30 [1.07, 1.47] \\
1/5 & Conduction & 415 & 0.79 [0.59, 1.07] \\
1/4 & Valence & 415 & 1.47 [1.28, 1.66] \\
1/4 & Conduction & 415 & 0.96 [0.72, 1.20] \\
1/3 & Valence & 184 & 1.79 [1.63, 1.97] \\
1/3 & Conduction & 415 & 1.22 [0.97, 1.48] \\
\addlinespace
\multicolumn{4}{@{}l}{\textit{Endpoint shifts at the one-fifth threshold ($n=412$)}}\\
\multicolumn{2}{@{}l}{Quantity} & Cells & Median [interquartile range] (eV) \\
\multicolumn{2}{@{}l}{Valence: anion minus neutral} & 412 & -0.009 [-0.073, +0.053] \\
\multicolumn{2}{@{}l}{Conduction: anion minus neutral} & 412 & +0.011 [-0.047, +0.106] \\
\multicolumn{2}{@{}l}{Valence: absolute shift $>0.3$ eV} & 8 & --- \\
\multicolumn{2}{@{}l}{Conduction: absolute shift $>0.3$ eV} & 25 & --- \\
\addlinespace
\multicolumn{4}{@{}l}{\textit{Neutral gaps ($n=415$)}}\\
\multicolumn{2}{@{}l}{HOMO--LUMO gap} & 415 & 3.37 [3.06, 3.69] \\
\multicolumn{2}{@{}l}{Mobility gap, one fifth} & 415 & 5.48 [5.36, 5.61] \\
\multicolumn{2}{@{}l}{Mobility gap, one quarter} & 415 & 5.82 [5.71, 5.96] \\
\end{longtable}
}

\clearpage
{\small
\setlength{\tabcolsep}{4pt}
\renewcommand{\arraystretch}{1.12}
\begin{longtable}{@{}lrrll@{}}
\caption{Si--N bond lengths and changes at the trap site. Values are medians [interquartile ranges]. Bonds are ranked by neutral length and followed by atom identity; $n$ is the number of contributing events (Fig.~S6a).}\label{tab:si_bonds}\\
\toprule
Route & Rank & $n$ & Neutral length (\AA) & $\Delta r$ (\AA) \\
\midrule
\endfirsthead
\caption[]{Si--N bond lengths and changes at the trap site. (continued)}\\
\toprule
Route & Rank & $n$ & Neutral length (\AA) & $\Delta r$ (\AA) \\
\midrule
\endhead
\midrule
\multicolumn{5}{r}{\emph{Continued on next page}}\\
\endfoot
\bottomrule
\endlastfoot
Polaron & 1 & 165 & 1.706 [1.683, 1.732] & +0.055 [+0.040, +0.069] \\
Polaron & 2 & 165 & 1.743 [1.719, 1.765] & +0.055 [+0.029, +0.080] \\
Polaron & 3 & 165 & 1.786 [1.766, 1.819] & +0.081 [+0.049, +0.147] \\
Polaron & 4 & 165 & 1.851 [1.816, 1.899] & +0.152 [+0.082, +0.272] \\
Induced K & 1 & 182 & 1.697 [1.679, 1.716] & +0.035 [+0.021, +0.052] \\
Induced K & 2 & 182 & 1.731 [1.714, 1.753] & +0.030 [+0.006, +0.047] \\
Induced K & 3 & 182 & 1.786 [1.749, 1.829] & +0.025 [-0.005, +0.068] \\
Induced K & 4 & 182 & 1.962 [1.909, 2.018] & +0.578 [+0.474, +0.746] \\
Native K & 1 & 65 & 1.653 [1.636, 1.667] & +0.073 [+0.061, +0.085] \\
Native K & 2 & 65 & 1.679 [1.661, 1.694] & +0.079 [+0.066, +0.095] \\
Native K & 3 & 65 & 1.705 [1.680, 1.720] & +0.087 [+0.074, +0.106] \\
\end{longtable}
}

\clearpage
{\small
\setlength{\tabcolsep}{4pt}
\renewcommand{\arraystretch}{1.12}
\begin{landscape}
\setlength{\LTcapwidth}{\linewidth}
\begin{longtable}{@{}p{0.14\linewidth}p{0.065\linewidth}p{0.09\linewidth}p{0.12\linewidth}p{0.09\linewidth}p{0.09\linewidth}p{0.085\linewidth}p{0.13\linewidth}p{0.07\linewidth}@{}}
\caption{N-centred events outside the three main trapping routes. Atom indices are one-based. Precursor and trap atoms identify the endpoint centres; their minimum-image separation is listed. The outside-region flag refers separately to the vertical-anion participation-region criterion in Methods.}\label{tab:si_unclassified}\\
\toprule
Cell & Anchor & Trap atom & Precursor atom & Outside region & Centre separation (\AA) & $\eta$ & $E_{\rm stab}^{\rm corr}$ (eV) & Trap state \\
\midrule
\endfirsthead
\caption[]{N-centred events outside the three main trapping routes. (continued)}\\
\toprule
Cell & Anchor & Trap atom & Precursor atom & Outside region & Centre separation (\AA) & $\eta$ & $E_{\rm stab}^{\rm corr}$ (eV) & Trap state \\
\midrule
\endhead
\midrule
\multicolumn{9}{r}{\emph{Continued on next page}}\\
\endfoot
\bottomrule
\endlastfoot
\texttt{box\_006} & N & 127 & 260 & yes & 9.3 & 0.205 & 2.40 & 638 \\
\texttt{box\_081\_r} & N & 32 & 208 & yes & 7.7 & 0.222 & 2.23 & 637 \\
\texttt{box\_170} & N & 72 & 228 & yes & 8.2 & 0.218 & 3.65 & 639 \\
\end{longtable}
\end{landscape}
}

\clearpage
{\small
\setlength{\tabcolsep}{4pt}
\renewcommand{\arraystretch}{1.12}
\begin{landscape}
\setlength{\LTcapwidth}{\linewidth}
\begin{longtable}{@{}>{\raggedright\arraybackslash}p{0.27\linewidth}>{\raggedright\arraybackslash}p{0.19\linewidth}p{0.16\linewidth}p{0.16\linewidth}p{0.16\linewidth}@{}}
\caption{Separation of trapping routes by structural fingerprints. Upper rows give pairwise areas under the receiver-operating-characteristic curve for single readouts and held-out logistic-regression scores of the 35-component fingerprint vectors. The longest neutral Si--N bond at the trap atom provides a reference for polaron--induced-K separation. Lower rows give the held-out fractions correctly classified from $(T^0,\eta)$ alone. Control rows repeat the comparison for the seven standardised channels of the trap atom without wavelets and for the readouts without the two coordination counts. Intervals are 95\% event-bootstrap intervals with scores and folds fixed; a zero-width interval at perfect separation reflects the fixed held-out scores, not zero uncertainty.}\label{tab:si_separation}\\
\toprule
Readout & Type & Polaron -- induced K & Polaron -- native K & Induced K -- native K \\
\midrule
\endfirsthead
\caption[]{Separation of the trapping routes by the structural fingerprint readouts. (continued)}\\
\toprule
Readout & Type & Polaron -- induced K & Polaron -- native K & Induced K -- native K \\
\midrule
\endhead
\midrule
\multicolumn{5}{r}{\emph{Continued on next page}}\\
\endfoot
\bottomrule
\endlastfoot
Neutral amplitude $T^0$ & single readout & 0.706 [0.651, 0.759] & 0.999 [0.997, 1.000] & 0.999 [0.996, 1.000] \\
Fingerprint change $\eta$ & single readout & 0.917 [0.884, 0.945] & 0.984 [0.968, 0.996] & 1.000 [1.000, 1.000] \\
Amplitude ratio $T^-/T^0$ & single readout & 0.659 [0.600, 0.717] & 0.964 [0.940, 0.985] & 0.998 [0.994, 1.000] \\
Precursor fingerprint $V^0$ & 35 components, held out & 0.833 [0.789, 0.874] & 1.000 [1.000, 1.000] & 1.000 [1.000, 1.000] \\
Trap fingerprint $V^-$ & 35 components, held out & 1.000 [1.000, 1.000] & 1.000 [1.000, 1.000] & 0.871 [0.822, 0.918] \\
Response $V^--V^0$ & 35 components, held out & 1.000 [1.000, 1.000] & 1.000 [1.000, 1.000] & 1.000 [1.000, 1.000] \\
$\eta$ without the coordination counts & control, single readout & 0.502 [0.437, 0.567] & 0.951 [0.921, 0.976] & 0.997 [0.991, 1.000] \\
Raw channels, precursor & control, held out & 0.882 [0.846, 0.917] & 1.000 [1.000, 1.000] & 1.000 [1.000, 1.000] \\
Raw channels, trap & control, held out & 1.000 [1.000, 1.000] & 1.000 [1.000, 1.000] & 0.900 [0.856, 0.937] \\
Raw channels, change & control, held out & 1.000 [1.000, 1.000] & 0.992 [0.981, 0.999] & 1.000 [1.000, 1.000] \\
Longest neutral Si--N bond at the trap atom & reference & 0.867 [0.831, 0.903] & --- & --- \\
\addlinespace
\multicolumn{5}{@{}l}{\textit{Held-out route assignment, share of events given their own route}}\\
 & All routes & Polaron & Induced K & Native K \\
Route assignment from $(T^0,\eta)$ (\%) & 95.7 [93.8, 97.5] & 89.5 [84.6, 93.9] & 100.0 [100.0, 100.0] & 99.7 [99.1, 100.0] \\
The same without the coordination counts (\%) & 72.8 [68.8, 76.7] & 61.5 [54.8, 68.0] & 75.7 [70.2, 81.1] & 93.5 [87.7, 98.2] \\
Most frequent route (\%) & 44.2 & --- & --- & --- \\
\end{longtable}
\end{landscape}
}

\clearpage
{\small
\setlength{\tabcolsep}{4pt}
\renewcommand{\arraystretch}{1.12}
\begin{landscape}
\setlength{\LTcapwidth}{\linewidth}
\begin{longtable}{@{}>{\raggedright\arraybackslash}p{0.43\linewidth}p{0.13\linewidth}p{0.39\linewidth}@{}}
\caption{Machine-readable source data accompanying the manuscript. The HSE Parquet package additionally contains the per-atom fields, fingerprints and physical-response tables used by the graph-wavelet analysis. Absolute energies and correction terms are supplied with the data.}\label{tab:si_data_manifest}\\
\toprule
File & Extent & Content \\
\midrule
\endfirsthead
\caption[]{Machine-readable source data accompanying the manuscript. (continued)}\\
\toprule
File & Extent & Content \\
\midrule
\endhead
\midrule
\multicolumn{3}{r}{\emph{Continued on next page}}\\
\endfoot
\bottomrule
\endlastfoot
\path{structures/q0/*.extxyz} & 455 files & relaxed neutral cells: cell matrix, coordinates, Mulliken charge and spin; header carries the vertical-leg correction and the precursor$\rightarrow$trap link \\
\path{structures/q-1/*.extxyz} & 455 files & relaxed anions at the same cell matrix; header carries the relaxed-leg correction and the trap$\rightarrow$precursor link \\
\path{energies/neutral_states_full.csv.gz} & 3,312,400 rows & all Kohn--Sham states of every neutral cell, both spins (3,640 per spin), raw and N~2s-aligned \\
\path{energies/charge_-1_states_full.csv.gz} & 3,312,400 rows & the same for the relaxed anion \\
\path{energies/charge_-1_vertical_states_full.csv.gz} & 3,312,400 rows & the same for the anion at the neutral geometry \\
\path{energies/*_states_wide.csv} & 91,910 rows each & states 591--691, the window of the IPR files \\
\path{energies/alignment.csv} & 455 rows & N~2s reference, alignment shift, HOMO/LUMO, source of each spectrum \\
\path{ipr/neutral_ipr_wide.csv}, \path{ipr/charge_-1_ipr_wide.csv} & 91,910 / 91,102 rows & atomic IPR per state, states 591--691; the anion table is in the state order of the spectrum and keeps the export index in \texttt{state\_export} \\
\path{charge_corrections.csv} & 455 rows & every correction scheme for both legs, raw total energies, $E_{\rm stab}^{\rm corr}$, trap and precursor identification, and the analysis set with the reason a pair is set aside \\
\path{cell_id_map.csv} & 455 rows & cell key of this deposit $\leftrightarrow$ structure id of the companion structure deposit \\
\path{MANIFEST.csv}, \path{CHECKSUMS.sha256} & 910 rows / 923 files & per-file metadata and checksums \\
\path{SI_tables/*.csv} &  & machine-readable versions of every table in this document \\
\end{longtable}
\end{landscape}
}